\documentclass[a4paper,11pt]{article}
\pdfoutput=1 

\usepackage{jheppub} 

\usepackage[T1]{fontenc} 
\usepackage{multicol}
\usepackage{multirow}
\usepackage{graphicx}
\usepackage{subcaption}
\usepackage{lineno}
\usepackage{xcolor}
\usepackage{natbib}

\title{\boldmath  Radioactivity control strategy for the JUNO detector}

\author[5]{Angel Abusleme}
\author[45]{Thomas Adam}
\author[66]{Shakeel Ahmad}
\author[66]{Rizwan Ahmed}
\author[55]{Sebastiano Aiello}
\author[66]{Muhammad Akram}
\author[29]{Fengpeng An}
\author[22]{Qi An}
\author[55]{Giuseppe Andronico}
\author[67]{Nikolay Anfimov}

% ...
\affiliation[5]{Pontificia Universidad Cat\'{o}lica de Chile, Santiago, Chile}
% ...
\affiliation[22]{University of Science and Technology of China, Hefei, China}
% ...
\affiliation[29]{East China University of Science and Technology, Shanghai, China}
% ...
\affiliation[45]{IPHC, Universit\'{e} de Strasbourg, CNRS/IN2P3, F-67037 Strasbourg, France}
% ...
\affiliation[55]{INFN Catania and Dipartimento di Fisica e Astronomia dell Universit\`{a} di Catania, Catania, Italy}
% ...
\affiliation[66]{Pakistan Institute of Nuclear Science and Technology, Islamabad, Pakistan}
\affiliation[67]{Joint Institute for Nuclear Research, Dubna, Russia}
% ...

\def\pbduz{$^{210}$Pb}

\def\poduo{$^{218}$Po}
\def\poduq{$^{214}$Po}
\def\udto{$^{238}$U}
\def\thdtd{$^{232}$Th}

\def\rnddd{$^{222}$Rn}
\def\kqz{$^{40}$K}

\def\Rlab{$r_{LS}$}
\def\Eth{$E_{th}$}
\def\Dpd{D$_{p-d}$}

\def\DT{$\Delta T_{p-d}$}

\def\be{\begin{equation}}
\def\ee{\end{equation}}

\abstract{ 
JUNO is a massive liquid scintillator detector with a primary scientific goal of determining the neutrino mass ordering by studying the oscillated anti-neutrino flux coming from two nuclear power plants at 53\,km distance.
The expected  signal anti-neutrino interaction rate is only 60 counts per day (cpd), therefore a careful control of the background sources due to radioactivity is critical. In particular, natural radioactivity present in all materials and in the environment represents a serious issue that could impair the sensitivity of the experiment if appropriate countermeasures were not foreseen. 
In this paper we discuss the background reduction strategies undertaken by the JUNO collaboration to reduce at minimum the impact of natural radioactivity. We describe our efforts for an optimized experimental design, a careful material screening and accurate detector production handling, 
and a constant control of the expected results through a meticulous Monte Carlo simulation program. We show that all these actions should allow us to keep the background count rate safely below the target value of 10\,Hz (i.e. $\sim$1 cpd accidental background)  in the default fiducial volume, above an energy threshold of  0.7~MeV.

}

\begin{document} 
\maketitle
\flushbottom

\section{Introduction}
\label{sec:intro}

The Jiangmen Underground Neutrino Observatory (JUNO)~\cite{yellow,ppnp} is a multipurpose experiment primarily designed to determine the neutrino mass ordering and precisely measure the neutrino oscillation parameters by detecting reactor anti-neutrinos. 
It is being built in the south of China at about 53\,km distance from the Yangjiang and Taishan nuclear power plants,  to allow the contemporary study of the solar and atmospheric neutrino oscillation sectors. 
The large detector volume --- 20,000 tons of liquid scintillator (LS) --- and the unprecedented energy resolution of 3\% at 1\,MeV~\cite{juno-calibration}, make JUNO  the largest LS-based, underground, neutrino observatory, capable of addressing many important topics in astro-particle physics. The extensive physics program of JUNO comprises supernova neutrinos, atmospheric neutrinos, solar neutrinos, and geoneutrinos, as well as new physics searches. A comprehensive discussion can be found in Refs.~\cite{yellow,ppnp}. 

The main detection channel for reactor anti-neutrinos in JUNO is the inverse beta decay (IBD) reaction on free protons (see Section\,\ref{sec:bkg-sources}). Because of its very low cross-section, only about 60 anti-neutrinos per day will be detected by JUNO, for a total available thermal power of 26.6\,GW$_{\textrm{th}}$. Therefore, strict control of radioactive  background is mandatory.

The JUNO detector is located in an underground laboratory with 700 m overburden, i.e. 1800~m.w.e. At this depth, the muon flux at JUNO site is of about 0.004 Hz/m$^{2}$ with a mean energy of 207 GeV~\cite{ppnp}. Hence a water Cherenkov detector is placed around the LS in order to detect and reject the muons. The rate of muons passing through the ultrapure water is of 10 Hz, while the rate of muons passing through the LS is of 3.6 Hz. Fast neutrons produced by muons passing through the rock and the detector materials may reach the LS and mimic an IBD event: the efficient muon tagging by the veto system of JUNO will enable removing most of these events leading to a low impact on the background budget of the fast neutrons. Muons and muon showers interact with $^{12}$C in the LS producing Z$\leq$6 isotopes by hadronic or electromagnetic processes: the produced $\beta$–n decaying nuclides 
can also mimic an IBD signal. The resulted $^9$Li/$^8$He are the most dangerous correlated background nuclides. 
In this respect, there are various physics-driven models for veto strategies to reduce the impact of the cosmogenic background in the different JUNO physics channels.
Geoneutrinos, produced by $^{232}$Th and $^{238}$U radioactive decay chains inside the Earth, are both a background for reactor anti-neutrinos due to the same interaction channel (IBD) and one of the JUNO physics cases. All these background sources (cosmogenics and geoneutrinos), which are related to experimental site location, are discussed elsewhere~\cite{yellow,ppnp}.

Natural radioactivity exists in all materials and represents a dominant source of background, which must be reduced with huge efforts on material selection and environmental control.
To reach JUNO sensitivity requirement for the neutrino mass ordering determination, it is essential to maintain the background count rate --- the so-called \emph{singles} rate --- due to the natural radioactivity 
below 10\,Hz~\cite{yellow}.
In this paper, we will discuss the natural radioactivity background reduction strategy pursued by JUNO to achieve this goal. 
The JUNO detector is described in Section~\ref{sec:setup}.
Section~\ref{sec:bkg-sources} details the natural radioactivity background sources and the anti-neutrino detection channel. The methodology for background reduction --- material screening and Monte Carlo simulations --- is presented in Section~\ref{sec:methodology}. Finally, the results are discussed in Section~\ref{sec:mc-results}. 

\section{The JUNO detector}
\label{sec:setup}

\begin{figure}[tbp]
 \centering 
 \includegraphics[width=1.0\textwidth]{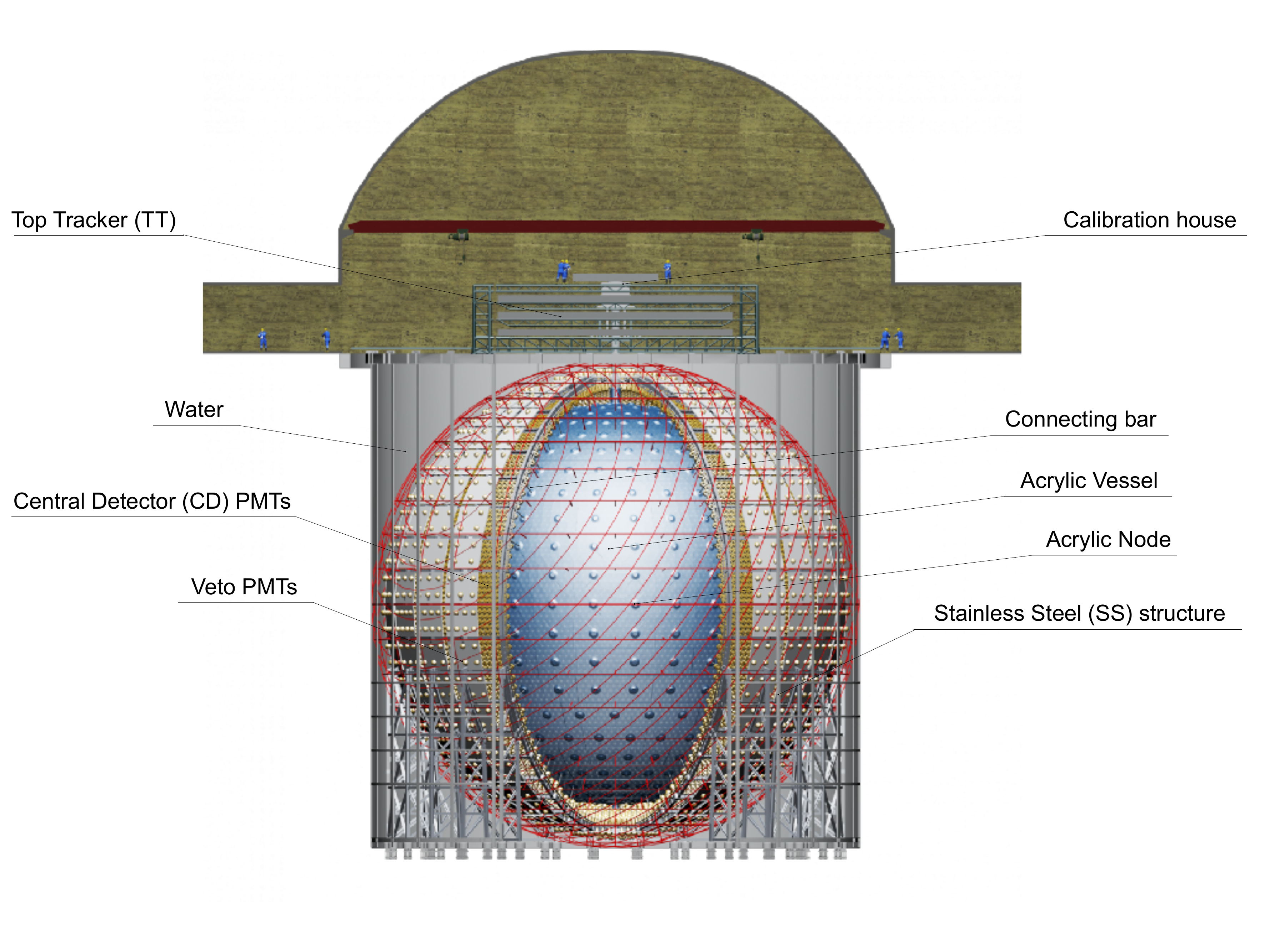}
  \hfill
  \caption{\label{fig:juno-schematic} Schematic drawing of the JUNO detector showing the main components of the experimental \cite{ppnp}.}
 \end{figure}
\begin{figure}[tbp]
 \centering 
  \includegraphics[height=7cm]{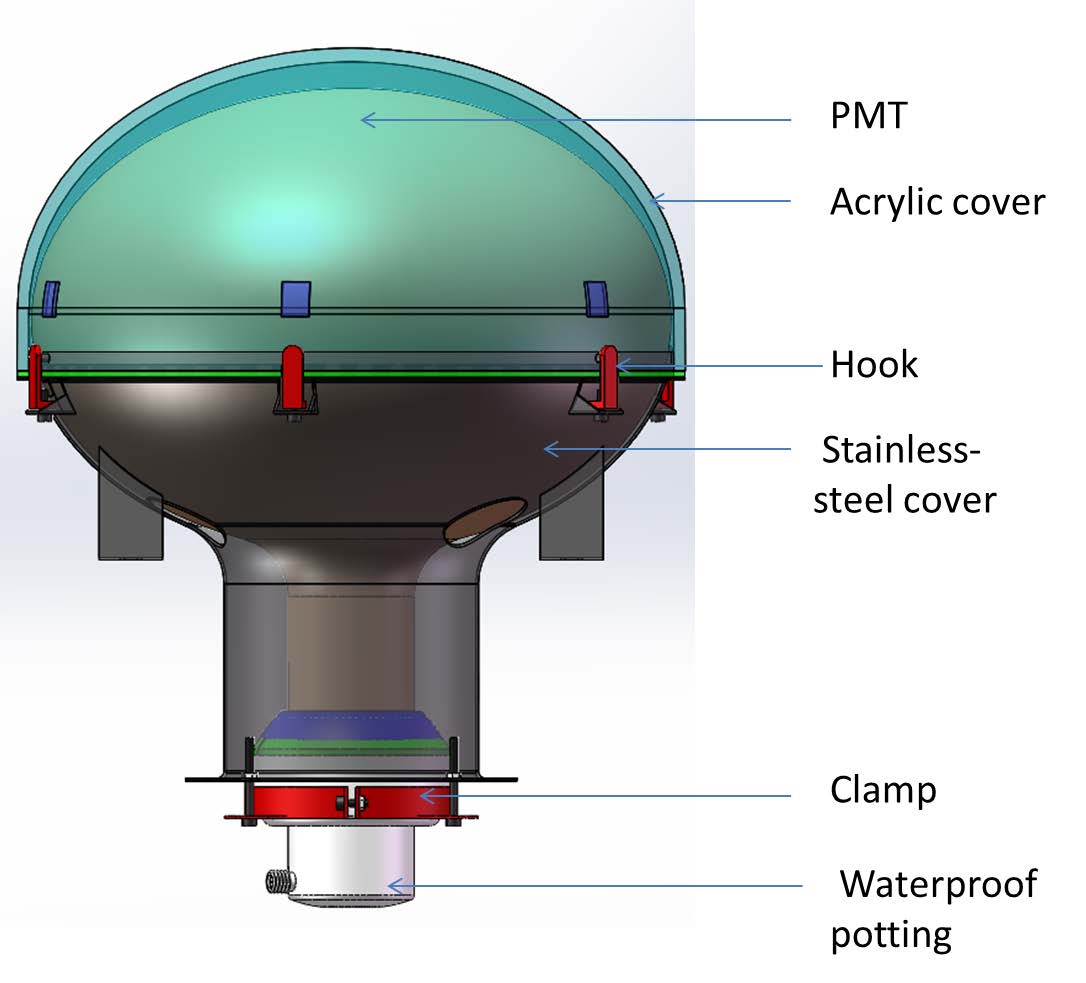}
\hfill
 \includegraphics[height=7cm]{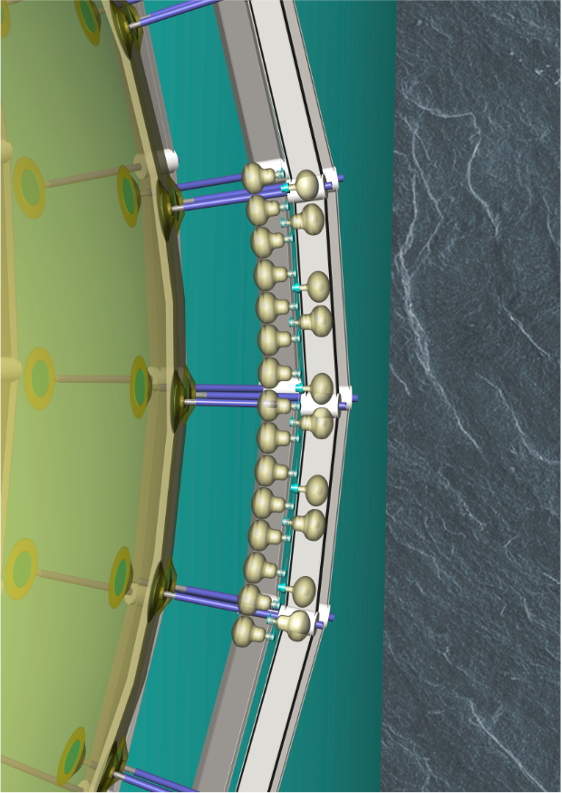}
 \caption{\label{fig:pmt-truss} Left: Schematics of an instrumented NNVT photomultiplier. Right: The acrylic vessel (yellow-green) anchored to the SS truss (light-gray) where the PMTs are installed and immersed in water (teal). The water pool (dark-gray -- not drawn to scale) has a minimum thickness of 3.9~m from the acrylic vessel in order to provide a sufficient shielding against the radioactivity of the rock. The inner part (2.3~m thick) is designed in order to shield the LS from the radioactivity of the PMTs. }
\end{figure}

The highly demanding performance requested to JUNO in terms of high statistical precision, excellent energy resolution ($\sim$3\%/$\sqrt{(E(\rm{MeV})}$), and low 
background have driven the dimensions and design of the experimental setup. It consists of a Central Detector (CD) and a water Cherenkov detector (WCD) laid in a water pool, and a muon tracker placed on top of the pool (indicated as Top Tracker, TT), as shown in Figure\,\ref{fig:juno-schematic}. 
In the CD, the huge LS mass -- 20\,kt  -- is contained inside a spherical acrylic vessel   with inner diameter of 35.40\,m and thickness of 12\,cm, for a total mass of about 580\,t of acrylic. 
The LS target has a room temperature density of 0.86\,g/mL and consists of Linear Alkyl Benzene (LAB) solvent, mixed with 2.5\,g/L of PPO (2,5-dyphenyloxazole) as fluor and 3\,mg/L bis-MSB (1,4-bis(2-methylstyryl) benzene) as a wavelength shifter~\cite{juno-scint}. 
The vessel is supported by a spherical stainless steel (SS) structure  with inner diameter of 40.1\,m, sitting on 30 pairs of SS legs safely rooted to the concrete floor of the water pool.
The anchoring of the acrylic vessel to the SS truss is ensured by 590 stainless steel rods (SS bars), which end at the vessel side with hinged connections within acrylic nodes to ensure the required stress relief.
The scintillation light produced by energy depositions in the LS volume is read by 17,612 20-inch photomultiplier tubes (LPMTs, for large PMTs) and 25,600 3-inch photomultiplier tubes (SPMTs, for small PMTs), which are installed on the inner side of the SS truss. The LPMTs are produced by two different companies: there are 5,000 20-inch dynode photomultipliers from Hamamatsu Photonics\,\cite{hamamatsu}, and 12,612 20-inch microchannel plate photomultipliers by Northern Night Vision Technology (NNVT) \cite{nnvt}.
All LPMTs feature a special protection  in case of implosion: 
$\sim$10\,mm thick acrylic semisphere on the top, supported by a 
$\sim$2\,mm thick stainless steel  semisphere on the bottom (left of Figure\,\ref{fig:pmt-truss}). 
The SPMTs are produced by HZC Photonics\,\cite{hzc}.
The PMTs' electronics is composed of two parts: the ``wet'' electronics is located few meters from the PMTs inside custom made stainless steel under water boxes (UWBs), while the ``dry'' electronics is placed in a dedicated room above the pool. Each UWB contains the customized high voltage (HV) module, front-end board and readout card for 3 LPMTs or 128 SPMTs.
The HV divider circuit to provide the working voltage to the PMTs is placed at the back of each phototube  inside a customized experimental volume with a waterproof potting. The cables connecting the PMTs to the UWBs and the UWBs to the ``dry'' electronics are kept inside waterproof SS bellows. 

The energy calibration is obtained by a redundant system of multiple sources (both radioactive and laser-based ones) and multidimensional scan systems, some of which will be inserted inside the vessel through an acrylic chimney. Calibration sources of different types and energies can be moved inside the LS volume 
or pulled inside a guide tube which surrounds the outside of the acrylic vessel and runs in a longitudinal loop. Details of the calibration systems and strategy are reported in Ref.~\cite{juno-calibration}. 
For the purpose of this paper, the calibration components that will be permanently mounted in the CD are the SS cables (with teflon anchors) used to insert the sources, the Teflon guide tube (with SS cables and sensors), and the ultrasonic sensor system receivers (which include Teflon, nickel, epoxy, piezoceramics, and copper) used to reconstruct the exact source positions within the detector.

The entire CD is submerged in a cylindrical water pool with a diameter of 43.5\,m and a height of 44.0\,m, providing sufficient water thickness in all directions (at least 3.9 m) to shield the detector from the radioactivity of the surrounding rock \cite{Li_2016}.
The inner part of the water pool, with 2.3\,m of water buffer, shields the LS from the radioactivity of the PMTs (right of Figure\,\ref{fig:pmt-truss}) with a 1.4\,m distance between acrylic vessel and the front face of the PMT glass bulb. 
To prevent  $^{222}$Rn (called also radon or Rn hereafter) diffusion from the external rocks from dissolving into the water, 5\,mm thick high-density polyethylene (HDPE) panels (liner) are sealing the pool walls and are sustained by a concrete barrier with a minimum thickness of 70\,cm placed in front of the cavity rock.
The water pool is equipped with about 2,400 LPMTs from NNVT to act as a WCD to veto cosmic muons. The CD and the WCD are optically separated with Tyvek. The WCD LPMTs are installed on the outside of the SS truss. Tyvek reflective foils provide a coating for the pool walls and the SS structure to increase the light collection efficiency. Finally, on the top of the water pool, a TT is installed to precisely measure the muon directions and support the veto strategies. It is composed by scintillating strips decommissioned from the Target Tracker of the OPERA experiment\,\cite{OPERA}, as described in Ref.\,\cite{JUNO-CDR}.

\section{Signal and radioactivity background}
\label{sec:bkg-sources} 

\subsection{Anti-neutrino signal and selection cuts} 
\label{sec:nu-cuts} 
JUNO detects electron anti-neutrinos via the IBD reaction, $\overline{\nu}_{e}+p\rightarrow e^{+}+n$. Here
the positron quickly deposits its kinetic energy and annihilates into two 511\,keV $\gamma$-rays, producing the prompt signal. The neutron, instead, scatters in the detector for tens of centimeters in the LS until it is thermalized and then captured by a free proton, with the subsequent release of a 2.2\,MeV $\gamma$-ray, which constitutes the delayed signal (mean capture time $\sim$200\,$\mu$s).
The energy deposited by the positron is related to the one carried by the anti-neutrino with a shift down of about 0.8\,MeV. Reactor anti-neutrino energies typically extend up to about 10\,MeV: thus, taking into account the IBD threshold energy of 1.8\,MeV, the deposited energy spectrum for reactor anti-neutrino analysis is spread between about 1 and 9\,MeV.

The IBD reaction shows a peculiar correlation in energy, time and space between the prompt and delayed signals. 
Therefore, the picking of anti-neutrino events is performed by applying, among others, a basic set of preliminary selection criteria (default cuts)~\cite{yellow}:

\begin{itemize}
     \item prompt signal energy: 0.7 MeV < $E_ p$ < 12 MeV;
     \item delayed signal energy: 1.9 MeV < $E_d$ < 2.5 MeV;
     \item time difference between prompt and delayed signal: 
     $\Delta T_{p-d}$ < 1.0 ms;
     \item distance between prompt and delayed signal:  $D_{p-d}$ < 1.5 m.
\end{itemize}

On the other hand, the accidental background from natural radioactivity (see Section~\ref{sec:natural-sources}) satisfies the above selection criteria in the same energy range, and cannot be discriminated by means of further software-based cuts.

The expected accidental coincidence rate $R_{acc}$ caused by natural radioactivity contaminants can be calculated as
\begin{equation}
 \label{sec5:eq1}
R_{acc}=R_p\cdot R_d\cdot\Delta T_{p-d}\cdot\epsilon
 \end{equation}
where $R_p$ and $R_d$ represent the prompt-like and delayed-like signal rates.
$\epsilon$ is an efficiency parameter that allows a further event selection by requiring a specific distance \Dpd\ between the position of occurrence of the prompt and delayed events within the detector. $\epsilon$ can be derived only by means of Monte Carlo simulations.
We optimize $R_{acc}$ with the help of a toy Monte Carlo, and tune  both the time and vertex correlation cuts between prompt and delayed signals according to the ratio of signal to background (as discussed later in Section~\ref{sec:bkg-impact}).

Therefore, despite the clear correlations in energy, time and position between the prompt and delayed signals, it is mandatory to maintain the natural radioactivity  background at extremely low levels. This can be achieved by a strict reduction strategy during the detector design and construction.

The measurement of solar neutrinos, on the other hand, fully rely on the detection of their elastic scattering off electrons in the detector medium.
Here only a fraction of the neutrino energy is transferred to the electron and the resulting electron recoil spectra are continuous even for incident mono-energetic neutrinos. The lack of a coincidence signature, in this case, poses even more stringent demands on the background level in the energy region below about 20\,MeV and was already studied in Ref.~\cite{JUNO-solar}.

\subsection{Natural radioactivity} 
\label{sec:natural-sources}

Natural radioactivity comes from all materials and can only be reduced by strict requirements on the material screening and environmental control. The water shielding around the central detector is effective not only for the fast neutrons generated by muons, but also for the radioactivity from the rock of the JUNO site. 
With a deposited energy up to 5~MeV which overlaps with the IBD energy spectrum, the radioactivity of the materials is one of the main sources of accidental background. 
The main contaminants  are the following:
\begin{itemize}
    \item natural long-lived radionuclides $^{238}$U and $^{232}$Th (with their decay chains supposed to be at secular equilibrium) and $^{40}$K;            
    \item natural medium-lived radionuclides $^{226}$Ra,  $^{210}$Pb/$^{210}$Bi, $^{210}$Po when secular equilibrium is broken in the \udto\ chain;
    \item natural gaseous radionuclide $^{222}$Rn;
    \item anthropogenic radionuclide $^{60}$Co.
\end{itemize}


All of these contaminants can be present in the various materials of the JUNO detector and may contribute to the singles rate measured in the  CD due to the energy deposition following their radioactive decays. It can correspond either to the prompt or the delayed event of the IBD reaction depending on their  energy and relative time. The expected IBD rate in the JUNO CD  induced by reactor neutrinos is about 60 counts per day~(cpd), while the singles rate from natural radioactivity should be controlled to less than 10 counts per second, leading to $\sim$1~cpd accidental coincidence (see Section~\ref{sec:bkg-impact}) with default anti-neutrino selection cuts, similar to $^{8}$He/$^9$Li and geoneutrinos background sources. However, for solar neutrino detection whose signal is a single event, this limit will be several orders of magnitude lower and achieved by applying more stringent fiducial volume (FV) and timing cuts to remove natural radioactivity and cosmogenic backgrounds, as described in a dedicated paper \cite{JUNO-solar}. We should point out that $^{14}$C and $^{85}$Kr are not considered here due to their decay energy (Q$_{\beta}$=156~keV and Q$_{\beta}$=687~keV, respectively) below the default energy cut for IBD reactions (0.7~MeV -- introduced in Section~\ref{sec:nu-cuts}). On the other hand, they are a very important background for low energy solar neutrino studies in JUNO.  

The radioactivity from external materials can also be effectively removed by software-based FV and energy threshold cuts, which can be both optimized based on the Monte Carlo simulation (see Section \ref{sec:methodology} and
 \ref{sec:mc-results}). Besides that, it is important to distinguish the {\it internal background}, coming from the LS itself and the {\it external background}, coming from the other parts of the JUNO detector. Indeed, for the internal background, all the radionuclides especially from the U/Th chains will contribute to the deposited energy whatever the nature of the emitted particles ($\alpha$, $\beta$, and $\gamma$) because the energy is directly released in the sensitive  volume. Assuming secular equilibrium, each of the radionuclides will contribute equally to the singles rate 
 (without energy cut considerations). In this case, a FV cut  would be useless for removing the background events, since they are uniformly distributed in the LS. 
 On the contrary, only a part of the radionuclides will contribute to the external background  considering the very short ranges of the $\alpha$ and $\beta$ particles in solid materials, which will exclude pure $\alpha$ and $\beta$ emitters. Only high energy $\gamma$s will be able to reach the LS and deposit energy, mainly coming from
$^{214}$Bi (1.76 and 2.20~MeV in $^{238}$U chain), $^{208}$Tl (2.61~MeV in $^{232}$Th chain), $^{40}$K  (1.46~MeV) and $^{60}$Co (1.17 and 1.33~MeV).

Radon coming from $^{238}$U chain is also a problematic gaseous radionuclide contributing to both internal and external backgrounds. Before  insertion in the JUNO vessel, the LS may be contaminated during its lifetime (production, transportation, and storage) with the radon in the air or emanated from the surrounding materials. As a result, an additional contribution from the long-lived isotopes $^{210}$Pb and its sub-chain $^{210}$Bi/$^{210}$Po has to be considered for  the internal background. As a progenitor of the $^{214}$Bi radionuclide, it may also contribute to the external background if present in the inner water pool. 
Finally, ($\alpha$,n) reactions can occur in the  LS or in the surrounding materials due to the U/Th impurities: anyway the impact in singles rate is expected to be much smaller than the ones just described due to a relative low cross section.

In the remaining of the paper we will focus on the strategies to reduce the impact of natural radioactivity on JUNO not to impair the sensitivity for the neutrino mass ordering determination.

\section{Methodology for background control}
\label{sec:methodology}

A proper design of the experimental setup is of paramount importance for the containment of the radioactivity background. The final arrangement of the detector is a delicate balance between the engineering plan and the need for the lowest possible count rate due to the spurious events.
The trade-off is usually achieved by a careful selection of the materials to be used for the construction of the apparatus, according not only to their mechanical characteristics but also to their intrinsic radiopurity. By means of \emph{ad hoc} Monte Carlo simulation of the expected radioactivity background, the best geometrical layout is thus evolved as the result of the radioactivity screening of the various materials and their positioning within the experimental setup.

One important point to be underlined for low background experiments is that very often the sensitivity needed to validate raw materials is at the cutting-edge of available screening techniques: this difficulty implies that the approval of certain materials or of particular production and cleaning protocols requires by itself to conceive non trivial dedicated test facilities.
JUNO surely benefits from the experience of past and running neutrino and dark matter experiments using the same type of detector, to select the proper  materials and related cleaning procedures to achieve its goals: the challenge comes from the pushing of ultra--low background techniques 
to the largest experimental scale.

In the next sections we will illustrate the experimental techniques used for the material screening and the simulation codes developed for the evaluation of the JUNO background.

\subsection{Material  assay and measuring techniques}
\label{sec:material-selection}


\begin{table}[tbp]
\centering
\begin{tabular}{|l|c|c|c|c|c|c|c|}
\hline
\multirow{3}{*}{Material} & \multirow{3}{*}{Mass} & \multirow{3}{*}{Radius} & \multicolumn{5}{c|}{Target impurity concentration}  \\
\cline{4-8}
& & & $^{238}$U & $^{232}$Th & $^{40}$K & $^{210}$Pb/$^{222}$Rn & $^{60}$Co  \\
& [t] & [m] & [ppb] & [ppb] & [ppb] & [ppb]/[mBq/m$^3$]  & [mBq/kg]   \\ 
\hline
\hline
Liquid Scintillator &&&&&&&\\
LS-reactor & \multirow{2}{*}{20000} & \multirow{2}{*}{0--17.7}
& 10$^{-6}$ &  10$^{-6}$ & 10$^{-7}$ &  10$^{-13}$ ppb & \\
LS-solar  
&&& 10$^{-8}$ &  10$^{-8}$ & 10$^{-9}$ &  10$^{-15}$ ppb & \\
\hline
Acrylic vessel & 580  & 17.7--17.8 & 0.001 & 0.001  & 0.001 & &  \\
Acrylic nodes & 28.5  & 17.8-17.9 & 0.001 & 0.001  & 0.001 & &  \\
\hline
Calibration parts & 0.04 & & 1.5 & 4.5 & 0.02 &&\\
\hline
SS structure &&&&&&&\\
-- truss & 1000 & 20.0-20.5 & 1 & 3 & 0.2 &&20 \\
-- bars  & 65 & 17.9-20.0 & 0.2 & 0.6 &  0.02 && 1.5  \\
\hline
LPMT glass  &&&&&&&\\
-- NNVT & 84.5 & 19.2-19.8 & 200  & 120 &  4 &&  \\
-- Hamamatsu & 33.5 & 19.2-19.8 &400 &400 &40 & &\\
-- Veto (NNVT) & 16.0 & 20.2-20.8 & 200  & 120 &  4 &&  \\
\hline
LPMT cover  &&&&&&&\\
-- acrylic & 110 & 19.2-19.4 & 0.003  & 0.01 &  0.01 &&  \\
-- SS & 150 & 19.4-19.8 & 0.4 & 2.5 &  0.12 & &\\
\hline
LPMT readout &&&&&&&\\ 
-- divider & 0.6 & 19.8-19.9 &3000 & 5000 & 100  &&\\
-- potting & 24.5 & 19.7-19.9 & 70 & 50 & 4  &&\\
-- UWB & 100 & 20.1-20.4 & 50 & 200 & 5 && 20\\
\hline
SPMT glass & 2.6 & 19.3-19.4 &400 &400 &200 & &\\
\hline
SPMT readout &&&&&&&\\ 
-- divider & 0.15 & 19.4 &3000 & 10000 & 200  &&\\
-- potting & 5.1 & 19.4-19.5 & 100 & 50 & 20  &&\\
-- UWB & 11 & 20.1-20.4 & 50 & 200 & 5 && 20\\
\hline
Water & 35000 & 17.8--21.8 & &&&10 mBq/m$^3$&    \\ 
\hline
Rock & && 10000 & 30000 & 5000 && \\
\hline
\end{tabular}
\caption{\label{tab:impurities}  Target values for the impurity concentrations in the different detector materials of the JUNO detector. For each detector component, the mass and its geometrical position, i.e. the radius quoted from the center of the LS volume, are reported. LS-reactor refers to the target impurity values required for mass hierarchy determination, whereas LS-solar refers to the target impurity values for solar neutrinos studies.}
\end{table}

A crucial prerequisite for the components of the setup of  ultra-low background experiments is the radiopurity  level. All materials must be selected according to their intrinsic low concentration of natural radioactive species, and any processing or handling must be carefully worked out in order not to accidentally contaminate the bulk or surface of the final product. This is achieved by setting up a radioactive screening program that includes different techniques to be exploited at the various stages of the component production according to the type of contaminant and to the  required sensitivity to reach the JUNO physics goals.  

Table\,\ref{tab:impurities} reports the minimal requirements for the radiopurity of the materials to be employed in the JUNO experimental setup. It is worth noting that the radiopurity requirements are less and less stringent for detector components with low mass and far from the vessel (from acrylic to SS truss), compared to the LS itself. A discussion on how these choices were finally made is postponed to Section\,\ref{sec:impurities}. The assumed values are expressed in Bq/kg or in mass concentration units~\footnote{For convenience,  the conversions between the two units used in this paper are reported:  1\,mBq of \thdtd\ activity per kg of material is equal to $2.5 \times 10^{-10}$\,g/g (or 0.25\,ppb) of \thdtd\ mass concentration in that material. Similarly, 1\,mBq/kg of \udto\ means $8.1 \times 10^{-11}$\,g/g (or 81\,ppt) of \udto, and 1\,mBq/kg of \kqz\ means $3.8 \times 10^{-12}$\,g/g (or 3.8\,ppt) of \kqz, corresponding to $3.2 \times 10^{-8}$\,g/g of natural K.} (g/g, i.e. grams of contaminant per gram of material, and its sub-multiples ppm\,$=10^{-6}$\,g/g, ppb\,$=10^{-9}$\,g/g, and ppt\,$=10^{-12}$\,g/g).
In the following we briefly describe the main measuring techniques available within the JUNO collaboration which were used to select the materials for the construction of the experiment according to the requirements of Table\,\ref{tab:impurities}. For some of them, e.g. the U/Th content in the LS, dedicated experimental facilities are being used to screen the samples at the desired sensitivity that are not described here. A comprehensive discussion of the different assays and measurement results, in fact, goes beyond the scope of the present paper. Few details can be found in Section\,\ref{sec:impurities}, where references to specific publications are given.

\paragraph{Low background spectroscopy}
Low background gamma spectroscopy with High Purity Germanium detectors (HPGe) is a common technique for screening materials needed  to build detectors for rare event studies. It allows a multi-radionuclide analysis of a sample in one measurement, giving access to its bulk activity for natural radioactivity (U and Th chains and $^{40}$K), and for cosmogenic or man-made gamma emitters ($^{60}$Co, $^{137}$Cs, etc) in the energy range  from 0 to 3\,MeV. The main advantage of this technique consists in measuring independently the activities of several radionuclides in the U/Th chains in order to check whether secular equilibrium is  achieved. It is of great importance e.g. for the Uranium chain since gamma spectroscopy is able to quantify $^{226}$Ra and $^{210}$Pb activities. Low background gamma spectroscopy with HPGe has typical sensitivities in the 10 ppt -- 10 ppb range or more (100 $\mu$Bq/kg -- 100 mBq/kg). Ultra-low background facilities in underground laboratories may reach the ppt scale ($\sim 10$\,$\mu$Bq/kg) with few tens of kg of sample.
Despite a moderate sensitivity compared to the other techniques described below (NAA and ICP-MS), it is the only technique able to measure short and medium-lived radioisotopes. The gamma spectrometers used for material screening in JUNO are HPGe detectors protected by passive and active shieldings and spread in several underground laboratories around the world (China JinPing underground Laboratory in China, Laboratoire Souterrain de Modane in France, Laboratori Nazionali del Gran Sasso in Italy\footnote{Courtesy of Dr.\,Matthias Laubenstein.}) or  sea level laboratories (IHEP in China, Milano-Bicocca in Italy, CENBG Bordeaux in France). They were used extensively to select a lot of components (such as SS from truss, bars and nodes, glass from LPMTs and SPMTs, electronics and calibration parts, etc.) or to investigate for the secular equilibrium break in some of them (acrylic, PPO, etc.).

\paragraph{Neutron Activation Analysis (NAA)}
NAA is a very sensitive method for qualitative and quantitative determination of trace elements based on the measurement of characteristic radiation from radionuclides formed by neutron irradiation of the material. 
The principle is very simple: following neutron capture by the  nuclide under investigation, a product radioactive nuclide is formed which is usually  $\beta$-unstable and decays  to excited states of the corresponding daughter nucleus, thus emitting characteristics $\gamma$-rays that can be measured by a HPGe spectrometer. NAA can achieve substantially greater sensitivity than direct $\gamma$-ray counting: it can be applied to quantify the concentration of natural contaminants (\udto, \thdtd, and \kqz) in detector materials that show no long-lived neutron activation products emitting  $\gamma$ lines which could interfere with the measurement.
The Milano-Bicocca group is being pursuing NAA on many materials since several years, using the TRIGA Mark II research reactor of University of Pavia (Italy) as the neutron source and the various HPGe detectors at the Radioactivity Laboratory of Milano-Bicocca University. Typical sensitivities are at ppt and sub-ppt levels \cite{NAA-talk}. For JUNO, NAA was used for the screening and quality control of acrylic, LAB, Teflon, and PPO.

\paragraph{Inductively coupled plasma mass spectrometry (ICP-MS)}
ICP-MS is widely used for screening materials of low background detectors due to its high sensitivity to trace $^{238}$U and $^{232}$Th. The ICP-MS located at IHEP in China is built in a Class 1000 clean room, and all other chemical operations are done in a Class 100 clean room. The count rate for ppt level of $^{238}$U/$^{232}$Th can reach $\sim$1000 counts per second (cps), and the detection limit can reach 0.01 ppt for pure water. For acrylic screening, a vaporization setup for acrylic pre-treatment is built in a Class 100 environment. With mature procedures for contamination control, the acrylic samples can be easily measured by ICP-MS to sub-ppt level in two days~\cite{sec4:ICP-MS_acrylic}. Besides that, this technique will play an important role on the quality control of the cleaning procedures and  the  purified water.

Laser Ablation ICP-MS is a complementary technique to ICP-MS to measure U/Th contaminations. The chemical preparation of the sample is replaced by a UV femtosecond laser used in an ablation mode. This promising technique has preliminarly achieved a surface or bulk sensitivity better than 10$^{-12}$ g/g level for U/Th in few minutes with only few tens of $\mu$g  sample \cite{LA-ICPMS} and is also well-suited to screen the surface treatment of the acrylic panels and other critical materials for JUNO.

\paragraph{Low background radon facilities}
There are multiple sources of radon that may contaminate the water or the nitrogen in JUNO: the contamination of the water or the nitrogen itself, the radon diffusion through barriers, the radon emanation from materials immersed in the  inner part of the water pool close to the acrylic sphere, etc. Thus, several facilities have been developed to reduce as much as possible the radon activity in the water pool or in the nitrogen gas. 

The radon measurement system developed at IHEP, which consists of an atomizer, a de-humidification system and an electrostatic radon detector, can be used to quantify the radon concentration in the water and gas~\cite{IHEPradon1,IHEPradon2}. The atomizer is a water vapor balancing device which could transfer the dissolved radon gas from water into air during the water flowing. The electrostatic radon detector determines the radon concentration by detecting the alphas from the \rnddd\ daughters  (\poduo\ and \poduq) with a Si-PIN photo-diode. The sensitivity of the radon detector is $\sim$\,5~mBq/m$^3$ for a one day measurement. The system will be used to measure and monitor the radon concentration in the ultra-pure water of the water Cherenkov detector  as well as in the  sealed nitrogen gas used on top of the JUNO pool.

A system to measure the radon activity of high purity nitrogen needed by JUNO was also developed~\cite{IHEPradon3}. The measurement setup contains two parts, a radon detection chamber (a 0.28~m$^{3}$ stainless steel tank) and a radon enrichment system (with  activated carbon as adsorbent). The background of radon detection chamber is $\sim$\,2~mBq/m$^{3}$ and the enrichment efficiency of the system was calculated to be about 50\% under 25~slpm flow  rate. A radon activity  at the level of 10~$\mu$Bq/m$^{3}$ could be measured by this system  in the high purity nitrogen. 

Low background radon emanation technique is complementary to the low background gamma spectroscopy technique and allows the measurement of the rate of radon atoms emanating from the surface of a given material depending on the activity of $^{226}$Ra, its long-lived progenitor. The emanation chamber used at CENBG laboratory in France is a vast stainless steel tank (0.7 m$^3$) allowing to fit in large material surfaces or a large number of samples. It is coupled to a low background electrostatic radon detector equipped with a Si-PIN diode \cite{Radon-detector} to perform the alpha spectroscopy of the $^{222}$Rn daughters. The typical sensitivity of the setup for 1 m$^2$ sample is few hundreds of atoms emanating per second and per square meter and the ultimate sensitivity can reach $\sim$ 10 \rnddd\ atoms/s/m$^2$ for a 30 m$^2$ sample \cite{Radon-emanation}. It is sensitive enough to screen large volume/surface materials that will be immersed in or around the JUNO water pool, such as rocks, 20-inch and 3-inch PMTs, HDPE liner, etc., in order to investigate the fulfillment of the \rnddd\ activity requirement in water (see Table~\ref{tab:impurities}).

A stainless steel bi-chamber was developed at CPPM laboratory in France to measure the radon transparency of the liner. The two identical chambers have a volume of $\sim$1.8~L. The sample to be measured is placed between the two chambers and the sealing is ensured by two flat Silicone gaskets. The upper chamber contains a \rnddd\ source  with a concentration of $\sim$\,740~kBq/m$^3$. The second chamber in which the radon passing through the sample is measured was initially filled with nitrogen. It is equipped with two valves and a second circulation pump, to homogenize the gas and fill a commercial 120 cm$^3$ Lucas cell used as a radon detector with a background of 15 Bq/m$^3$. The \rnddd\ transparency down to 2$\cdot$10$^{-5}$ can be measured by this setup for a material such as the HDPE liner.

\subsection{Monte Carlo simulations}
\label{sec:mc-simulations}

A whole detector simulation is realized to evaluate the radioactivity contributions in terms of singles rate from the main detector materials listed in Section~\ref{sec:setup}. The geometry of the experimental setup is reconstructed by the simulation code with highest possible detail. Each component is uniformly contaminated with the various sources described in Section~\ref{sec:bkg-sources} and its impact on the background count rate is then computed. The outcome of these simulations is an invaluable guidance during both the experimental design and the material selection processes, to choose the best solutions that reduce the impact of the dangerous background.
In the end, the results will be useful for the evaluation of the overall radioactivity background budget as well as for the optimization of both the FV  and the energy threshold cuts for the data analysis, depending on the physics channels under study. 

The official JUNO offline software is based on the SNiPER framework (described in Appendix\,\ref{sec:sniper}) and was developed according to the demanding requirements of the experiment profiting from the modularity and the flexibility offered by the SNiPER environment~\cite{sec4:sniper1,sec4:sniper2}. It comprises the physics generator, the detector simulation, and the electronics simulation as separated modules, which are integrated to return a complete reproduction of the real events. In the lack of an experimental benchmark while the experiment is under construction, two additional software codes, ARBY (described in Appendix\,\ref{sec:arby}) and G4-LA (described in Appendix\,\ref{sec:cenbg}), based on completely different logical architectures, were used for the simulation of the JUNO background, in order to have a validation of the background estimation outcomes.

\subsection{Validation of the Monte Carlo outputs}
\label{sec:mc-validation}

To validate the results of the JUNO background simulations, both energy spectra and overall count rates induced by natural radioactivity contaminants (Section\,\ref{sec:natural-sources}), as obtained by the three codes, were compared.
For the validation process, the simulated geometry included just the acrylic vessel and the LS inside it: despite the chosen experimental configuration was quite simple,  these studies involved all relevant physics phenomena and allowed to highlight unexpected bugs either in the physical implementation or in the logic of the simulations.
For a meaningful comparison, a common set of basic physical hypotheses was adopted for the three codes: i) the chosen GEANT4 version was 10.04.p2~\cite{Geant4-web} with the \textit{Livermore} low energy electromagnetic physics list; ii) all codes finally used the Geant4 particle source primary generator for the properties related to each radioactive decay, i.e. daughter nucleus, mean life, decay modes, branching ratios, and  emission spectra; iii) from the experience of the Daya Bay experiment~\cite{DayaBay-scint}, which used a scintillation detector very similar to the JUNO one, the secondary particle production thresholds in Geant4 were set at 0.1\,mm for electrons and at 1\,mm for gammas; iv) in the propagation of protons, alpha particles and nuclear recoils, the Geant4 \textit{StepFunction} (the computation of the mean energy loss per propagation step) was set at default values; v) finally, the scintillation non-linearity was described by means of the generalized Birks semi-empirical formula. Here, the energy converted to scintillation photons, $E_{scint}$, is related to the stopping power $dE/dx$ of a charged particle of kinetic energy $E$ via:
\begin{equation}
E_{scint}=S\int_{0}^{E} {\frac{dE}{1+kB(\frac{dE}{dx})+C(\frac{dE}{dx})^2}}
\label{eq:birks-formula}
\end{equation}

\noindent with $kB$ and $C$ being the Birks coefficients. 
The parameter $S$ is a normalization factor which gives the scintillation efficiency. 
\begin{table}[tbp]
\centering
\begin{tabular}{|l|c|c|}
\hline
 & $kB/\rho \times 10^{3}$  & $C/\rho^2 \times 10^{6}$\\
Particle type & [g/cm$^2$/MeV] & [(g/cm$^2$/MeV)$^2]$ \\
\hline 
\hline
Electrons, Positrons & 6.5  & 1.5  \\
Alphas, Nuclear recoils & 3.705   & 1.5  \\
Protons & 6.5   & 1.5  \\
\hline
\end{tabular}
\caption{\label{tab:birks-coeff} Values of the Birks' coefficients of Equation\,\ref{eq:birks-formula} divided by the LS density $\rho$ ($\rho = 0.86$\,g/cm$^3$) used to compute the ionization quenching in JUNO simulations.}
\end{table}
In the Birks model, only a fraction of the energy deposited in the LS by the ionizing particle is actually transferred to the fluorescent molecule (2.5\,g/L of PPO for JUNO~\cite{juno-scint}) and ready to be converted in scintillation light, a fact that is usually referred to as \textit{ionization quenching} of the light yield. Table\,\ref{tab:birks-coeff} reports the Birks' coefficients used for the present JUNO simulations, inherited from Daya Bay experience~\cite{DayaBay-scint}.

For the purpose of the validation of the background simulations, only the energy deposited and stored in the LS -- $E_{scint}$ in Equation\,\ref{eq:birks-formula} -- was considered, i.e. nor optical propagation nor light readout were involved at this step with SNiPER. The Monte Carlo outputs of the three simulation codes were thus converted into deposited energy spectra in the range between 0 and 12\,MeV with 10\,keV binning and  without energy threshold applied. Nor detector time resolution nor energy resolution were included. The whole LS volume --- i.e. scintillator radius \Rlab\ equal to the acrylic sphere inner radius, \Rlab\,=\,17.7\,m, hereafter indicated also as DV (detector volume) --- was initially considered as the detecting medium (no fiducial volume cut applied) in order to count all energy depositions happening at any place within the vessel.

\begin{figure}[tbp]
\centering 
\includegraphics[width=1.0\textwidth]{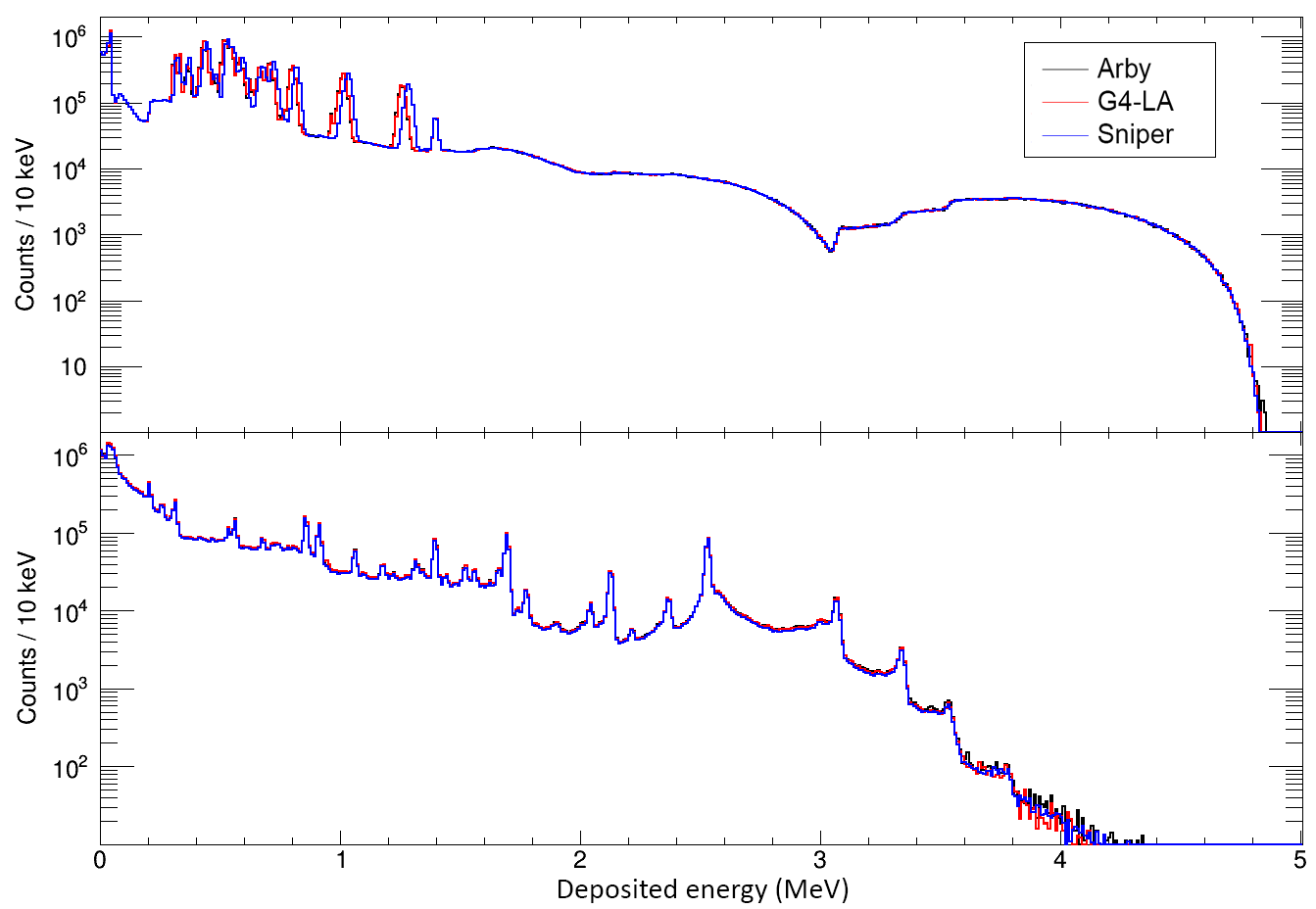} 
\caption{\label{fig:LS-acrylic-spectra} Superposition of the sum spectra obtained by considering all energy depositions in DV following \udto, \thdtd, \kqz, and \pbduz\ decays uniformly distributed within the LS (top) and the acrylic vessel (bottom), as simulated by each of the three Monte Carlo codes. For any radioactive species, $10^6$ ($10^7$) decays were considered, respectively. In the top plot, the black and red histograms are overlapping, while the blue one is slightly shifted at higher energy only for the quenched alpha peaks (see discussion in the text).}
\end{figure}

\begin{figure}[tbp]
\centering 
\includegraphics[width=.6\textwidth]{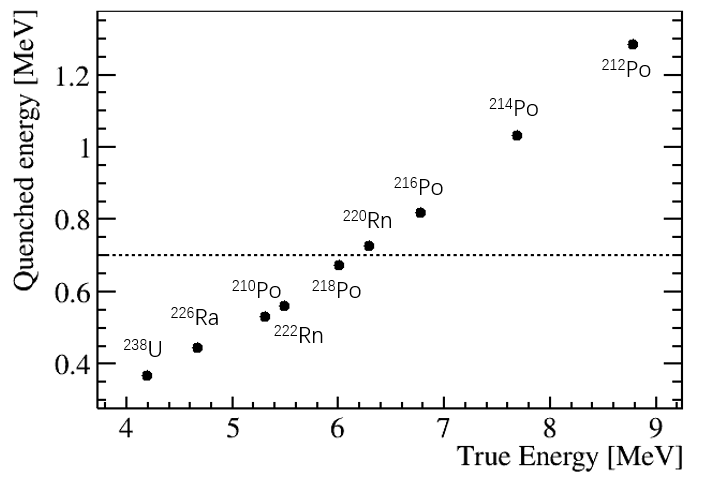} 
\caption{\label{fig:alphaQuenching-plot} The quenched alpha energy  resulting from the SNiPER simulation as a function of the true energy for the main alpha peaks of both \udto\ and \thdtd\ chains is shown as black points. The default 0.7 MeV energy threshold used for IBD event selection is shown with a dashed line.}
\end{figure}

The first set of simulations were devoted to the LS contaminants: in this case, the complete natural decay chains of \thdtd\ and \udto\ and the sub-chain starting with \pbduz\ (following the secular equilibrium break caused by \rnddd\ in the \udto\ chain), as well as the natural contaminant \kqz, were considered separately. For each radioisotope or decay chain progenitor, a total of $10^6$ nuclei were uniformly distributed inside the whole scintillator volume and the energy depositions following each decay sequence recorded for the comparative analysis.
The resulting summed energy spectra from the four contaminants obtained by each of the three software codes are superimposed in Figure\,\ref{fig:LS-acrylic-spectra} (top spectrum) in the energy region up to 5\,MeV, where the majority of the events are distributed. The alpha peak broadening visible in the spectra is due to the fluctuations of the  energy deposited (and consequently quenched) at each step along the particle track. As one can see, the features of the spectra are finally in very good agreement. The residual differences are attributed to the implementation of the quenching calculation, but they are not relevant for the discussion of the background impact in JUNO.
The position of all quenched $\alpha$ peaks is consistent within 25\,keV among the three codes: this spread has to be compared with the peak broadening due to step energy deposition fluctuations, of the order of $6-15$\,keV depending on the energy. We have also to note that the energy resolution of the detector (not included at this stage in the simulations) ranges from about 20 to 30\,keV in the quenched $\alpha$ peak energy interval. Figure\,\ref{fig:alphaQuenching-plot} shows the quenched alpha energy as a function of the true energy for the main alpha peaks of both \udto\ and \thdtd\ chains, as resulting from the SNiPER simulation. The quenching factor, i.e. the ratio between true and quenched energy, is ranging from 12 to 7 for true alpha energies going from 4 to 9 MeV. Most of the alpha events will be removed by the default 0.7 MeV energy threshold cut (see later) illustrated by the dashed line.
On Figure\,\ref{fig:LS-acrylic-spectra} (top spectrum), the end-point around 4.8 MeV is induced by the decay of the $^{208}$Tl (Q$_\beta$=5.0 MeV). The energy difference is due to the quenching of the electrons produced in LS after multiple Compton scattering of the gammas from $^{208}$Pb de-excitation.

For what concerns the acrylic vessel, the same \thdtd, \udto, or \kqz\ contaminants were uniformly distributed within its bulk volume. A total of $10^7$ decays for each of the radionuclides were simulated by the three codes and the following energy depositions in DV recorded for the comparative analysis. The obtained results are summarized in Figure\,\ref{fig:LS-acrylic-spectra} (bottom spectrum). Also in this case the agreement among the three software is extremely good.
It has to be noted that the acrylic vessel simulation results in Figure\,\ref{fig:LS-acrylic-spectra} include also the contributions of the acrylic nodes and the acrylic chimney.

To convert the number of events collected by the detector into expected count rates, one has to make hypotheses on the concentration of the residual impurities in the material of interest. In the case of the components of the JUNO detector, these inputs are the target values reported in Table\,\ref{tab:impurities} and discussed later in Section\,\ref{sec:impurities}. 

\begin{figure}[tbp]
        \begin{subfigure}{0.5\textwidth}
            \centering
            \includegraphics[width=\textwidth]{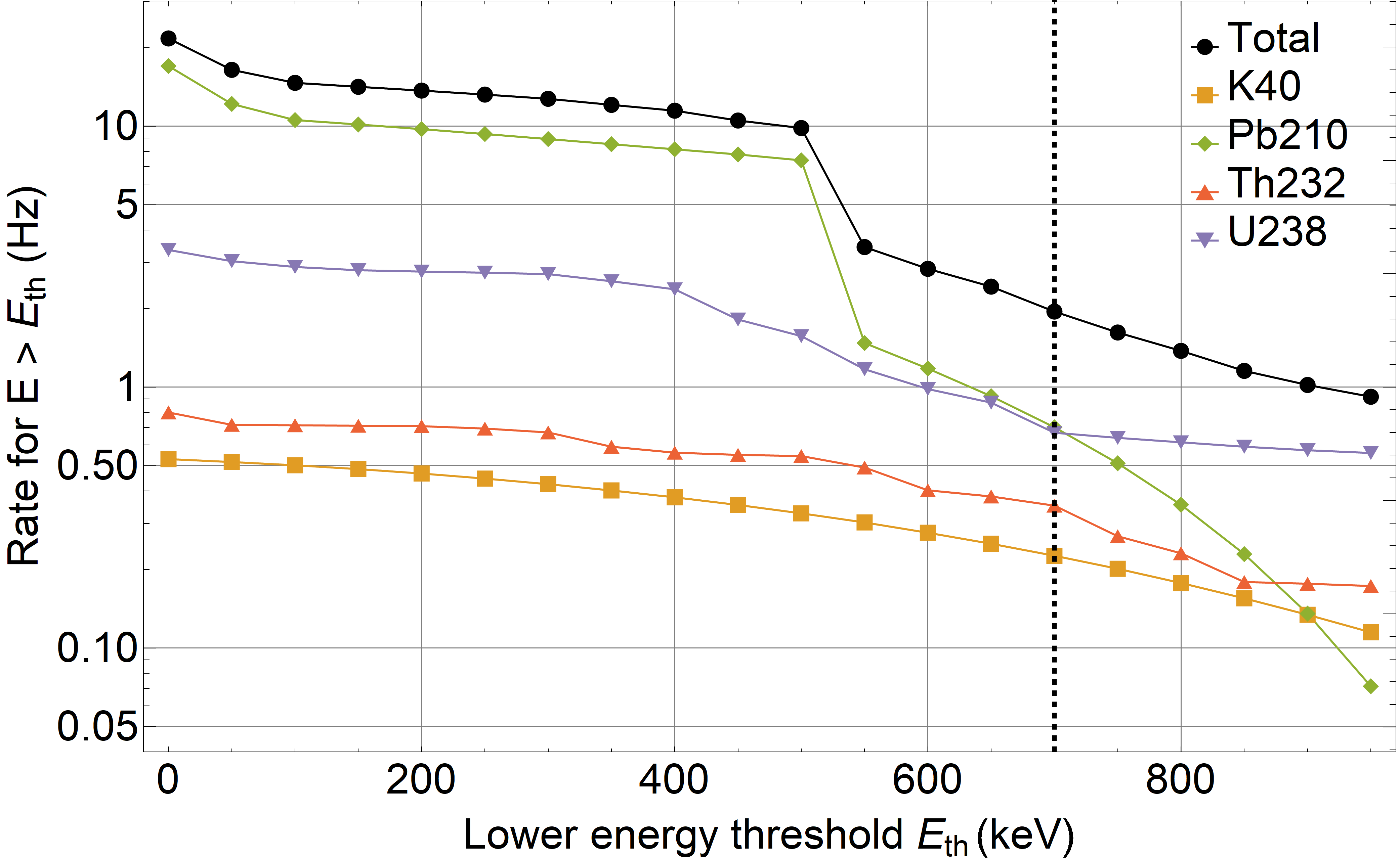}
            \caption[]{\label{fig:lab-Eth-plot} }
        \end{subfigure}
        \begin{subfigure}{0.5\textwidth}
            \centering
            \includegraphics[width=\textwidth]{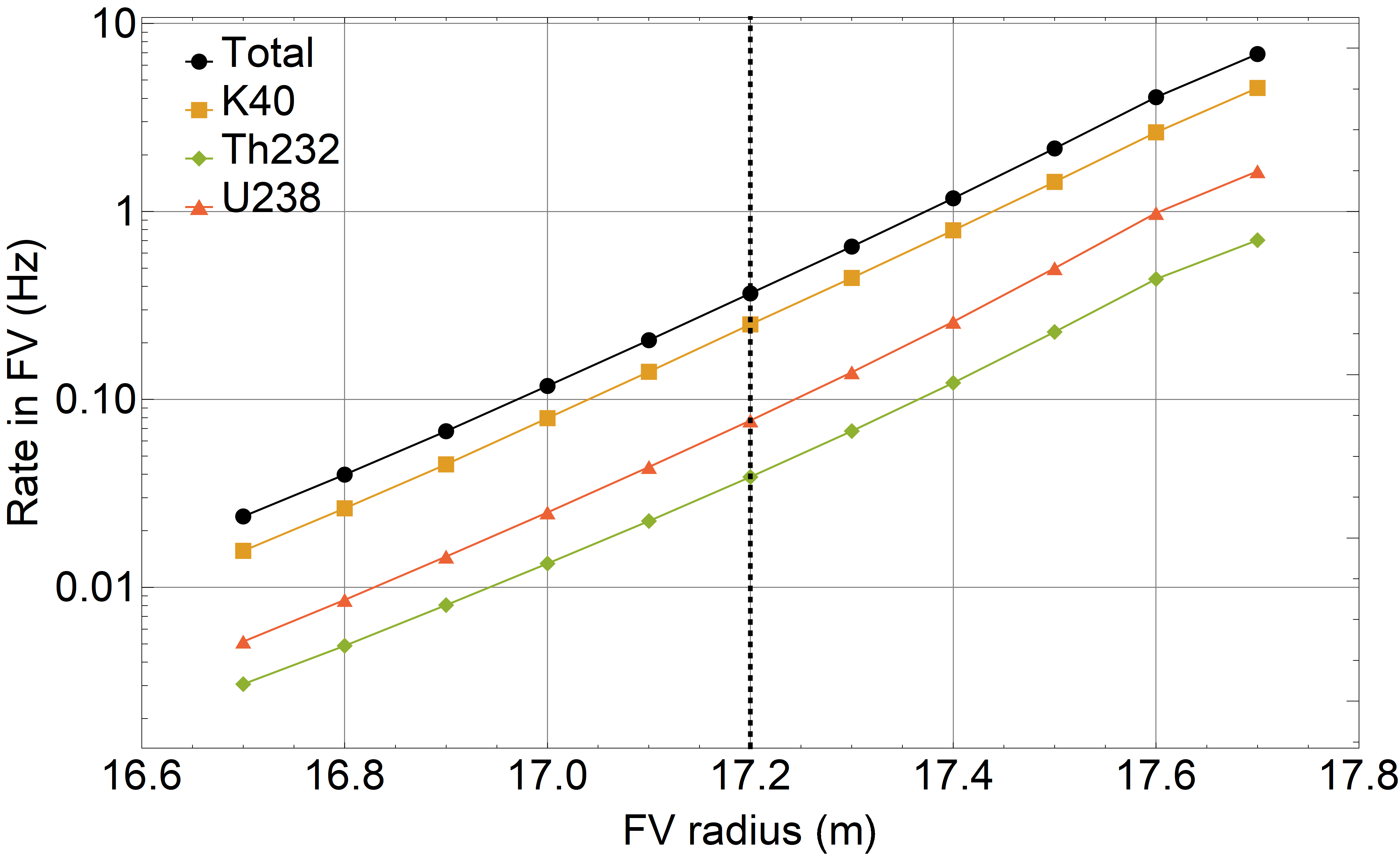}
            \caption[]{\label{fig:acrylic-FV-plot}}
        \end{subfigure}
        \caption[]{(a) Expected count rates in DV due to the residual contaminations of the LS (LS-reactor impurity inputs in Table\,\ref{tab:impurities}) as a function of the energy threshold \Eth\ applied to the simulated spectra. Solid lines  connect the points to guide the eye. \Eth\,=\,0.7 MeV is the threshold energy value chosen for the neutrino mass ordering analysis by JUNO (shown by the dashed line). (b) Expected background count rates for \Eth\,=\,0.7\,MeV due to uniform \kqz, \thdtd, and \udto\ contaminations in the bulk of the acrylic sphere as a function of different radius quotes \Rlab\ chosen to define the experimental FV. Solid lines  connect the points to guide the eye. Currently, the default JUNO FV for anti-neutrino analysis is spherical, with radius \Rlab\,=\,17.2\,m (shown by the dashed line).} 
        \label{spectra}
\end{figure}

The count rates generated in DV by the decays of the LS contaminants (LS-reactor inputs in Table\,\ref{tab:impurities}) uniformly distributed within the liquid scintillator, are shown in Figure\,\ref{fig:lab-Eth-plot} as a function of the energy threshold \Eth\ applied to the spectrum of each simulated contaminant. As one can see, the resulting rates decrease abruptly as the chosen threshold goes beyond the energy of the quenched alpha peaks, especially those belonging to the \udto/\pbduz\ chains.
For IBD analysis, a good compromise is an energy threshold of 0.7\,MeV, well below the prompt energy of the positron (E$_{e^+} \geq 1$\,MeV) and high enough to reduce the accidental rate to an acceptable value that does not impair the mass ordering sensitivity. This choice of the energy threshold for the neutrino mass ordering analysis is the default value assumed by JUNO at this stage, as anticipated for the $E_p$ energy selection criteria in Section~\ref{sec:nu-cuts}, and will be supposed in all later discussions in this paper. 

The count rates induced in DV by  the LS-reactor and the acrylic vessel impurities (from Table\,\ref{tab:impurities}) were calculated for the three codes above the default energy threshold  for comparison. The results are reported in Table~\ref{tab:singles-noFV}, separately for each nuclide. The relative uncertainty on the global rates among the three codes is smaller than 2.5\%, both for the LS and the acrylic simulations.

\subsection{Comparative impact of material contaminations}

The next step after the validation process was to evaluate the impact on the background rate of the residual impurities contained in all other materials of the JUNO detector with SNiPER and ARBY tools. Also in this case, the contaminants were uniformly distributed within the material of interest and the Monte Carlo codes were set to simulate the decay processes up to the energy depositions in the LS volume. The results are reported in Table\,\ref{tab:singles-noFV} as count rates in the energy region between 0.7\,MeV and 12\,MeV in DV, and are listed separately for the considered contaminant. 
The achieved agreement between the two simulation codes is within 7\% on all the considered components of the JUNO setup, and the achieved statistical uncertainty is less than 1\%.
The impurity inputs used to derive the contributed rates are the target values of Table\,\ref{tab:impurities}. Apart from LS and acrylic, the number of simulated decays for each radionuclide was 10$^8$ for all materials: the only exception was the rock of the underground laboratory, for which $2\times 10^{10}$ decays were considered for each of the contaminants. In this case, a hollow cylinder of internal radius equal to 21.75\,m and total height of 43.5\,m (i.e. the pool dimensions) was uniformly contaminated with \udto, \thdtd, or \kqz\ inside a thickness of 20\,cm (higher thickness values would make the simulations extremely inefficient due to self-absorption). The chosen material for the cylinder was SiO$_2$ with a density  close to 2.6\,g/cm$^3$, the average density of the rock at the experimental site. Despite the long simulation times, no events were detected for \udto\ and \kqz\ in the scintillator volume, so only upper limits at 90\% C.L. are listed in Table\,\ref{tab:singles-noFV} for the background induced by those contaminants in the laboratory rock. Given the geometrical characteristics and presumed contaminations of the rock simulation, these results can be considered as proxies for other materials at similar distances from the detector center, like the liner covering the rock walls placed in front of the concrete barrier or the veto detector placed on top of the pool. 

\begin{table}[tbp]
\centering
\begin{tabular}{|l|c|c|c|c|c|}
\hline
\multirow{2}{*}{Material}  & \multicolumn{5}{c|}{Count rates in DV (deposited energy) -- \Eth\ = 0.7\,MeV }  \\
\cline{2-6}
& {$^{238}$U} [Hz] & {$^{232}$Th} [Hz] & {$^{40}$K} [Hz] & {$^{210}$Pb} [Hz] & {$^{60}$Co} [Hz]\\
\cline{2-6}
\hline
\hline
LS-reactor & 0.70 & 0.37  & 0.23  & 0.69  & \\ 
\hline
Acrylic & 1.74 & 0.74 & 4.65 & &  \\ 
\hline
Calibration parts \hspace{5mm} & 0.5  &0.6 &0.2 &&\\  
\hline
SS structure &&&&&\\
-- truss &  0.02  & 0.08   & 0.002 & & 0.01 \\ 
-- bars & 3.6 & 4.8  & 1.0 & &4.3 \\ 
\hline
 LPMT glass &&&&&\\ 
 -- NNVT  & 6.04  & 4.03 & 0.17 & & \\ 
 -- Hamamatsu &4.79 & 5.33 & 0.69 && \\ 
 -- Veto (NNVT) &0.17  & 0.14  &0.003 &&\\ 
\hline
LPMT cover &&&&&\\ 
-- acrylic & 0.0001  & 0.0004 & 0.0006 &&\\ 
-- SS & 0.01  & 0.08  & 0.004 &&\\ 
\hline
LPMT readout &&&&&\\ 
-- divider  & 0.10  & 0.23 & 0.004 &&\\
-- potting   & 0.11 & 0.11  & 0.007 &&\\
-- UWB  & 0.21 & 1.2 & 0.02 & &0.003\\
\hline
SPMT glass  & 0.24  & 0.30 & 0.21 &&\\ 
\hline
SPMT readout &&&&&\\ 
-- divider  &  0.08 &   0.37  &   0.009 &&\\
-- potting  &   0.09 &  0.06 &  0.03 &&\\
-- UWB  & 0.03 & 0.16  &  0.002 & &$4\times 10^{-4}$\\
\hline
 Water (\rnddd) &0.77$^{\dagger}$ & &&&\\ 
\hline
 Rock   & <\,0.01$^{\ddagger}$  &0.05& <\,0.09$^{\ddagger}$ &&\\
\hline
\end{tabular}
\caption{\label{tab:singles-noFV} Background count rates in detector volume (DV) in the energy range between 0.7\,MeV and 12\,MeV induced by the various detector components with the contaminant concentrations  of Table\,\ref{tab:impurities}. 
The results are given for each contaminant separately, for a complete picture. Only the deposited energy is considered in these results, i.e. no optical propagation nor detector energy resolution were applied.
 \\
$^{\dagger}$This simulation refers to
a uniform contamination of 10\,mBq/m$^3$ of \rnddd.\\
$^{\ddagger}$Upper limits at 90\%\,C.L.
}
\end{table}

Concerning \rnddd\ distributed in the water of the pool, the simulation was performed considering only the inner water pool with a water shell of  2.3\,m thickness placed at a radius of 17.8\,m from the detector center: this choice allows to attain reasonable computation times and is justified by the fact that this layer is the closest to the CD. 

The “node and bar” simulation was performed considering just the stainless steel components incorporated in the nodes at the acrylic vessel side  and the stainless steel bars connecting the truss and the acrylic sphere, since the acrylic material making the nodes is much cleaner than the stainless steel and its contribution is already included in the results of the vessel. The outcomes are reported under the SS bars entry in Table~\ref{tab:singles-noFV}.
 For the calibration system, all components that are permanently mounted on the CD (described in Section~\ref{sec:setup}) were simulated separately and the results added together in Table\,\ref{tab:singles-noFV}; the contribution from the calibration sources (permanently stored in the calibration house on top of the detector) is found to be negligible.
For the LPMTs, the contributions of the radioactivity of the glass and of the protection cover are reported as  independent entries.
Moreover, given the different impurity concentration inputs for Hamamatsu and NNVT PMTs or the different positions within the setup of the PMTs (e.g. veto PMTs with respect to ``signal'' PMTs), the glass contributions for the different LPMTs are listed separately.
Finally, the signal readout simulations for both the LPMT and the SPMT systems were done separately for the components sitting on the back of the photomultipliers (divider and potting) and for the electronics boards placed in the under water boxes (UWBs) \cite{ppnp}.

Despite the assumed impurity level in the various components reported in Table\,\ref{tab:impurities} are among the smallest achievable by available techniques, the overall expected background count rate in Table\,\ref{tab:singles-noFV} is quite significant. The only means to reduce it is to shield the external sources by applying a software-based FV cut. Figure\,\ref{fig:acrylic-FV-plot} shows how effective is the rate reduction with increasing thickness of the external ``dead layer'' in the LS volume in the case of the background induced by the acrylic sphere contaminants (the closest to the detector FV). At this stage, the ``dead layer'' thickness chosen for IBD analysis by the JUNO collaboration is 50\,cm, which reduces the count rate by about one order of magnitude. Therefore, in the remaining of the paper a spherical FV with radius \Rlab\,=\,17.2\,m will be considered as the default value.

\section{Results of the Monte Carlo simulations}
\label{sec:mc-results}

In this last part of the paper we discuss the outcomes of the background Monte Carlo simulations performed with the SNiPER software after enabling the full reproduction of the event formation, i.e. taking into account all physical and geometrical effects that come into play following the energy deposition in the LS. Therefore, in this section all results include the energy resolution, the optical propagation, the charge conversion,  and the non-uniformity response of the detector (these effects were deeply described in previous works\,\cite{juno-calibration,DayaBay-scint}).  

The assumed impurities for the main components of the JUNO detector, which are the inputs to compute the expected background impact, are first analysed in Section~\ref{sec:impurities-intro}. The simulation outcomes are then reported in Section~\ref{sec:bkg-rates} in the chosen energy interval between 0.7\,MeV and 
12\,MeV.  
Finally, the different possible scenarios resulting from the accidental coincidences that could resemble an IBD event are
discussed in Section~\ref{sec:bkg-impact}.

\subsection{Target impurity concentrations in the JUNO detector and environment}
\label{sec:impurities-intro}

Table\,\ref{tab:impurities} reports the assumed upper limits for the concentration of the natural contaminants in the materials used for the JUNO detector. They should be regarded as the target values that the JUNO collaboration has set for the expected activities due to the radioactive impurities in the detector components: some of those values are based on preliminary measurements of the final products (acrylic, SS structure, PMT glass, components of calibration system), others are derived from the literature or from the experience of similar experiments. The motivations for those choices are given in Section~\ref{sec:impurities}.

Special attention must be devoted to the measures implemented to keep the probability of environmental contamination as low as possible, in particular during the detector installation which represents a delicate phase where the risk of nullifying all previous efforts is particularly high. This is discussed in Section~\ref{sec:environment}.

\subsubsection{Radioactivity inputs for the detector components}
\label{sec:impurities}

The LS is the key component of the JUNO detector. The residual contaminations of Table\,\ref{tab:impurities} in the case of IBD analysis are assumed based on the experience of the Borexino\,\cite{Borexino} and KamLAND\,\cite{KamLAND} experiments.
JUNO LS will undergo a sophisticated purification procedure that foresees four steps: Al$_2$O$_3$ filtration column (to improve optical properties), distillation (to remove heavy metals and improve the transparency), water extraction (to remove radioisotopes from U/Th chains and \kqz), and steam stripping (to remove gaseous impurities, such as Kr and Rn). Preliminary tests performed with prototype plants indicated that those assumptions for \udto\ and \thdtd\ (LS-reactor inputs of Table\,\ref{tab:impurities}) are within reach, so they are taken as the reference radiopurity requirements for the neutrino mass ordering physics channel.
The OSIRIS (Online Scintillator Internal Radioactivity Investigation System) quality-check detector\,\cite{osiris}, which will be placed just before the LS injection point into the acrylic vessel, is designed to assure the fulfilment of the 10$^{-15}$\,g/g requirements for \udto\ and \thdtd\ via a real-time measurement before filling the detector.
The huge dimensions of JUNO required careful planning of the LS purification plants to correctly size them according to the unprecedented LS volume that will be processed. Additional key points of the plant design are severe \rnddd\ tightness requirements for all pipes (see Section\,\ref{sec:environment}), strict demands on the quality of the water (\thdtd\ and \udto\ $\leq 10^{-16}$\,g/g, \rnddd\ $\leq 10$\,mBq/m$^3$) provided to all purification stages by the ultra pure water system that will be installed in the laboratory, and heavy prerequisites for the quality of the pure N$_2$ (\rnddd\ $\sim 10~\mu$Bq/m$^3$) that will be supplied by the nitrogen system to the purification plants and the LS containers, including the water pool. Moreover, strict cleaning requirements are foreseen for all mechanical components of the purification plants as well as severe dust control during the production and the installation. A constant check of the achievements during the installation phase will be pursued by means of the OSIRIS detector and of \emph{ad hoc} laboratory tests exploiting NAA coupled to coincident $\beta-\gamma$ measurements. All these precautions will hopefully grant the LS-solar contamination assumptions of Table\,\ref{tab:impurities}. Once again, all these requirements represent challenges that were never faced before the JUNO experiment, since the unprecedented LS mass prevents from reiterating the purification steps after the detector filling.

The aims in terms of radiopurity of the acrylic vessel are very similar to those achieved by the SNO solar neutrino experiment~\cite{SNO}. The chosen production company proved to have the expertise and the proper facilities to sustain the production and assembly of the huge acrylic panels needed to build the JUNO sphere: a dedicated Class 10,000 clean room was set up at the company site, the  mold is cleaned with deionized water, all pipes and containers are carefully washed, and \rnddd\ concentration is constantly monitored in all work areas. After the final grinding and polishing processes, a thin protection film made of polyethylene (PE) will be placed on each acrylic panel as UV protection and to prevent \rnddd\ diffusion and dust deposition on the surfaces: the film will be removed as last step during the vessel installation at the experimental site (see Section\,\ref{sec:environment}). All production stages are performed under the strict supervision of the JUNO collaboration, to guarantee high quality optical properties and negligible radioactive contamination during all phases of the sphere construction. 
Preliminary measurements of the bulk radioactivity of the produced acrylic panels show that the limits reported in Table\,\ref{tab:impurities} are safely within reach. A severe control of the mass production is being pursued by means of NAA and ICP-MS. The residual contaminants on the acrylic surfaces is being monitored by LA-ICPMS. Gamma spectroscopy on random samples will also be performed profiting of the most sensitive HPGe detectors to confirm that no contamination out of secular equilibrium, in particular for the \udto\ chain, occurred during the panel production. 

The structure that sustains the acrylic vessel is entirely made of SS. More than 1\,kton of low radioactivity SS -- grade 304L -- is produced by smelting of the molten iron in a clean furnace; no scrap steel is used. Also in this case, the whole production process is performed under the supervision of the JUNO collaboration. From the background point of view, the most critical components of this structure are the SS bars used to anchor the acrylic vessel to the truss itself. In fact, those rods will be embedded in the vessel through the acrylic node (see Figures~\ref{fig:sfig21} and \ref{fig:sfig3}), therefore they will be placed very close to the detector volume. For this reason, the requirements for the radiopurity of the SS bars in Table\,\ref{tab:impurities} are about one order of magnitude more severe than the ones for the SS truss. Preliminary gamma spectroscopy measurements confirmed the good quality of the purchased SS. Radioactivity checks will be performed during the production process to confirm the compliance with the upper limits of Table\,\ref{tab:impurities}.

The other critical component for the JUNO background budget is the PMT glass. Table\,\ref{tab:impurities} details the agreed maximum impurity concentrations reported in the contracts signed with the PMT vendors. A huge R\&D effort was put by the NNVT company, in close cooperation with the JUNO  collaboration, into controlling all glass production steps to reach the subscribed limits~\cite{sec5:MCP}. Gamma spectroscopy measurements on few glass bulb specimens were also performed with different HPGe detectors to verify compliance of both Hamamatsu LPMT glass and HZC Photonics SPMT glass \cite{cao2021mass} with the impurity concentration requirements during the mass production. 

Concerning the $^{222}$Rn in water, the requirement of 10 mBq/m$^3$ in Table 1 is based on the experience from SuperKamiokande and SNO experiments \cite{Nakano_2020,Blevis_2004}. This is quite challenging because of the radioactivity of the rock at the JUNO site (10 ppm for \udto/$^{226}$Ra) and the high radon level expected in the environment. A prototype of the water purification system  demonstrated the ability to decrease the Rn in water below 10 mBq/m$^3$ by using degassers and a microbubble generator. During the data taking, the ultrapure water will continuously circulate at a flow of 100 tons/h from the purification system to the top and bottom inlets of the inner water pool towards the outer water pool. This strategy will prevent a possible loading of Rn in water when circulating near the pool edges, where the Rn activity is expected to be higher than at the center. To prevent the Rn diffusion from the rock and the concrete to the pool, a HDPE liner will be installed on the walls of the water pool. Preliminary measurements of its transparency performed in laboratory have demonstrated a sufficient Rn suppression factor by more than 4 orders of magnitude using HDPE samples with 5 mm thickness. In addition, the combination of a N$_2$ blanket (purified at 1 mBq/m$^3$ Rn level) in overpressure and a thick Tyvek cover on top of the pool will isolate the water from the air of the underground laboratory. The radon in the inner water pool may also come from the Rn emanation of the detector components in contact with the inner water pool, especially the PMT glass, the dominant material considering its radioactivity: several specimens from the three types of bare PMTs (NNVT, Hamamatsu, and HZC) were screened by means of the Rn emanation facility described in Section \ref{sec:material-selection} and have shown a negligible contribution compared to the 10 mBq/m$^3$ target activity. Some efforts have still to be done on $^{226}$Ra removal in water with resins. All these strategies and R\&D efforts are converging in order to fulfill the Rn requirement of Table~\ref{tab:impurities}, but cannot guarantee it regarding possible unexpected variation of the Rn level during JUNO data taking. Anyway, there are some margins and an increase of the Rn activity, for example, by a factor of 5 in the inner water pool would imply an increase of the count rate in the FV within 0.3 Hz, i.e. well below other contributions.
 
\subsubsection{Environmental contamination control}
\label{sec:environment}

Even though the selected materials for the detector are compliant with the JUNO requirements, they can be contaminated afterwards by the environment during production, transportation, installation, as well as during data taking. As a result, special protections and surface cleaning are essential for all the critical materials after their final installation in the detector and before filling the detector with ultra pure water and LS. 

Special attention must be devoted to the surface cleanliness of the tanks or the equipment that will be directly in contact with the LS during the purification steps. The baseline limit for the residual dust on the surface is 0.1 mg/m$^2$ assuming that the radioactivity in the dust is similar to the one of the rock (see Table\,\ref{tab:impurities}): in this way the resulting contamination should not exceed the 10$^{-16}$ g/g level in 20 kt of LS.

The cleanliness of the environment during installation and vessel filling is also very important, since the acrylic will be exposed to the air for a limited time after the thin PE film removal and subsequent final cleaning, and before the filling with the LS. Our baseline for the cleanliness of the air inside the acrylic sphere should reach a level better than Class 10000.

The radon concentration in the air of the underground laboratory can reach more than 100~Bq/m$^3$ with a risk of \rnddd\ to be dissolved in the LS, leading to the long-lived $^{210}$Pb radionuclide. It is thus very important to have stringent leakage requirements. The leakage test is planned  by means of vacuum force technology. Assuming the pressure differential to be 1~bar and the radon concentration in underground air equal to 100~Bq/m$^3$, the leakage for the single component should reach $\sim$4$\times$10$^{-6}$ mbar$\cdot$L/s with vacuum-air test. The amount of dust and rock that may remain in the water pool should be minimized by a strict quality control of the surfaces before filling the detector with ultrapure water. Details concerning the procedures of cleanliness and leakage control are still under finalization and will be presented in forthcoming JUNO technical papers.

\subsection{Expected background count rates in JUNO}
\label{sec:bkg-rates}

\begin{figure}[h]
 \centering 
 \includegraphics[width=0.7\textwidth]{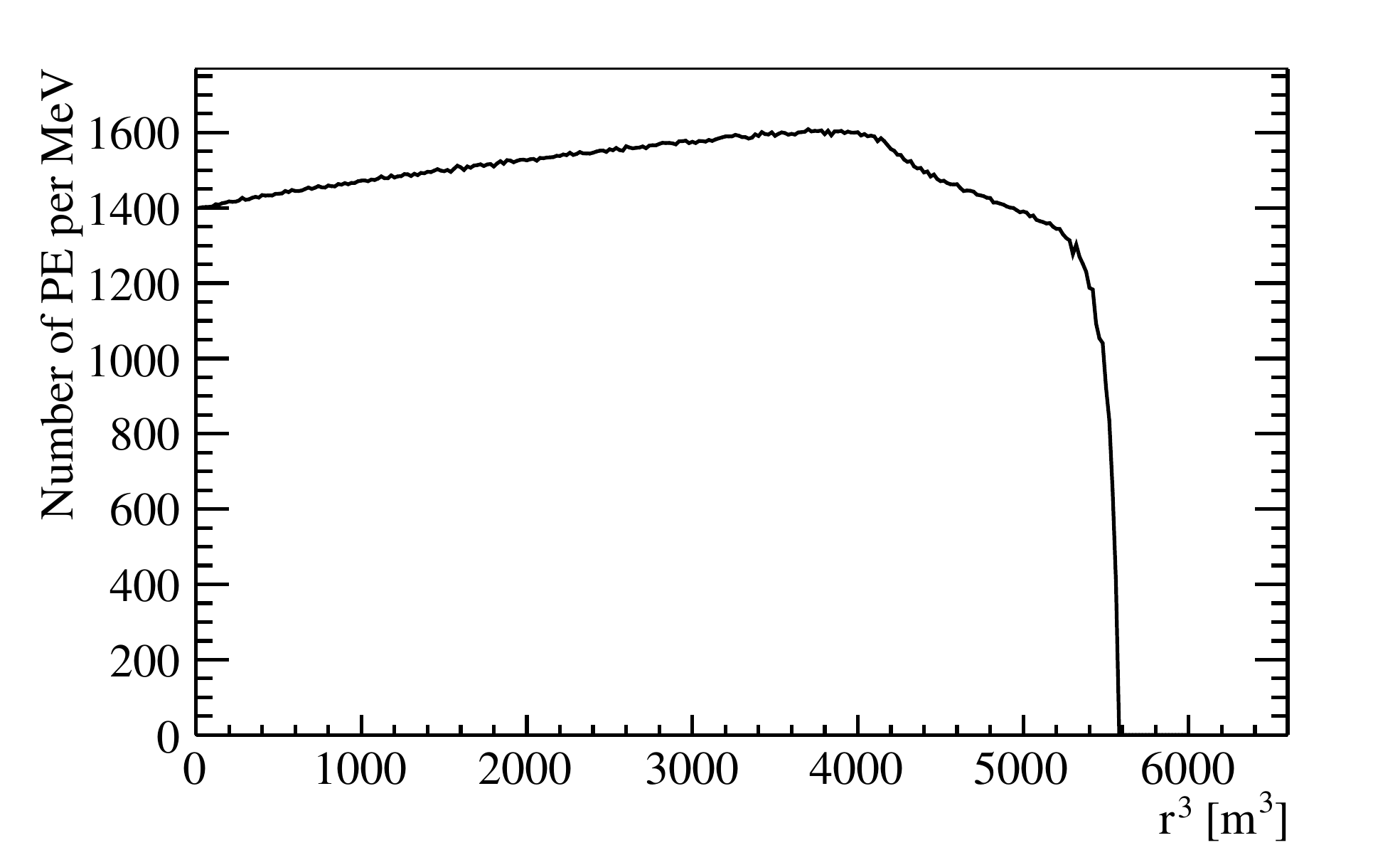}
  \hfill
  \caption{\label{fig:lightyield} Uniformly distributed 1 MeV gammas are simulated in the LS, and the total number of PEs from both LPMTs and SPMTs as a function of volume is shown in this figure.}
 \end{figure}

\begin{figure}[h]
 \centering 
 \includegraphics[width=0.9\textwidth]{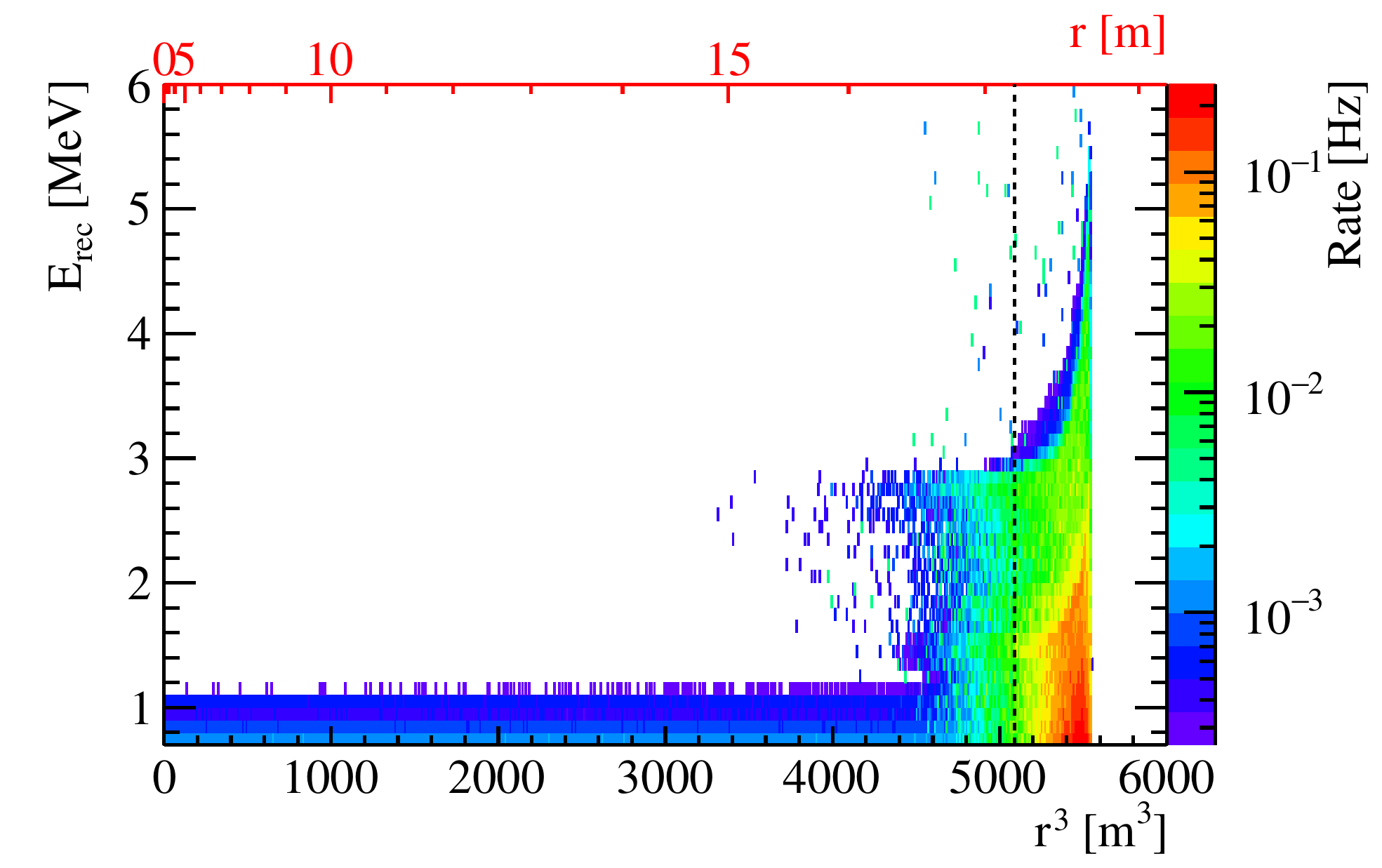}
  \hfill
  \caption{\label{sec5:singlesR3E} The singles event rate as a function of  volume (bottom X-axis) or radius (top X-axis) and reconstructed energy E$_{rec}$ (Y-axis). The default FV radius cut is shown as the dashed line.}
 \end{figure}

Full event simulation is performed with the SNiPER framework, including  the energy resolution, the optical propagation, the charge conversion,  and the non-uniformity of the detector response. The radioactivity coming from the main materials are simulated with the same statistics presented in Section~\ref{sec:mc-simulations}, and the total number of  photoelectrons (PEs) collected by both LPMTs and SPMTs is evaluated for each event. 
After the energy conversion, the obtained spectra of the reconstructed energy $E_{rec}$ are analysed to derive all needed information.  The conversion from the total number of PEs to the energy unit (MeV) is carried out by simulating the uniformly distributed 1 MeV gammas depositing energy in the whole LS volume. 
The energy non-linearity of the detector response is out of the scope of this paper (see Ref.~\cite{juno-calibration} for details) and thus a single energy value is sufficient to derive the reconstructed energy. 
The resulting total number of PEs as a function of volume is shown in Figure~\ref{fig:lightyield}, with $\sim$1400 PEs expected at the center of the detector. The sharp decrease  of the number of PEs at large radii is due to the energy leakage near the edge (Compton scattering) and total reflection of the optical photons at the interface between the acrylic and water, which will lead to larger statistical fluctuations on the reconstructed energy. Combining this effect with the energy resolution, this explains why the reconstructed energy from natural radioactivity can be spread up to 6 MeV (see later Figure~\ref{sec5:singlesR3E}).	    

\begin{figure}[tbp]
        \begin{subfigure}{0.5\textwidth}
            \centering
            \includegraphics[width=\textwidth]{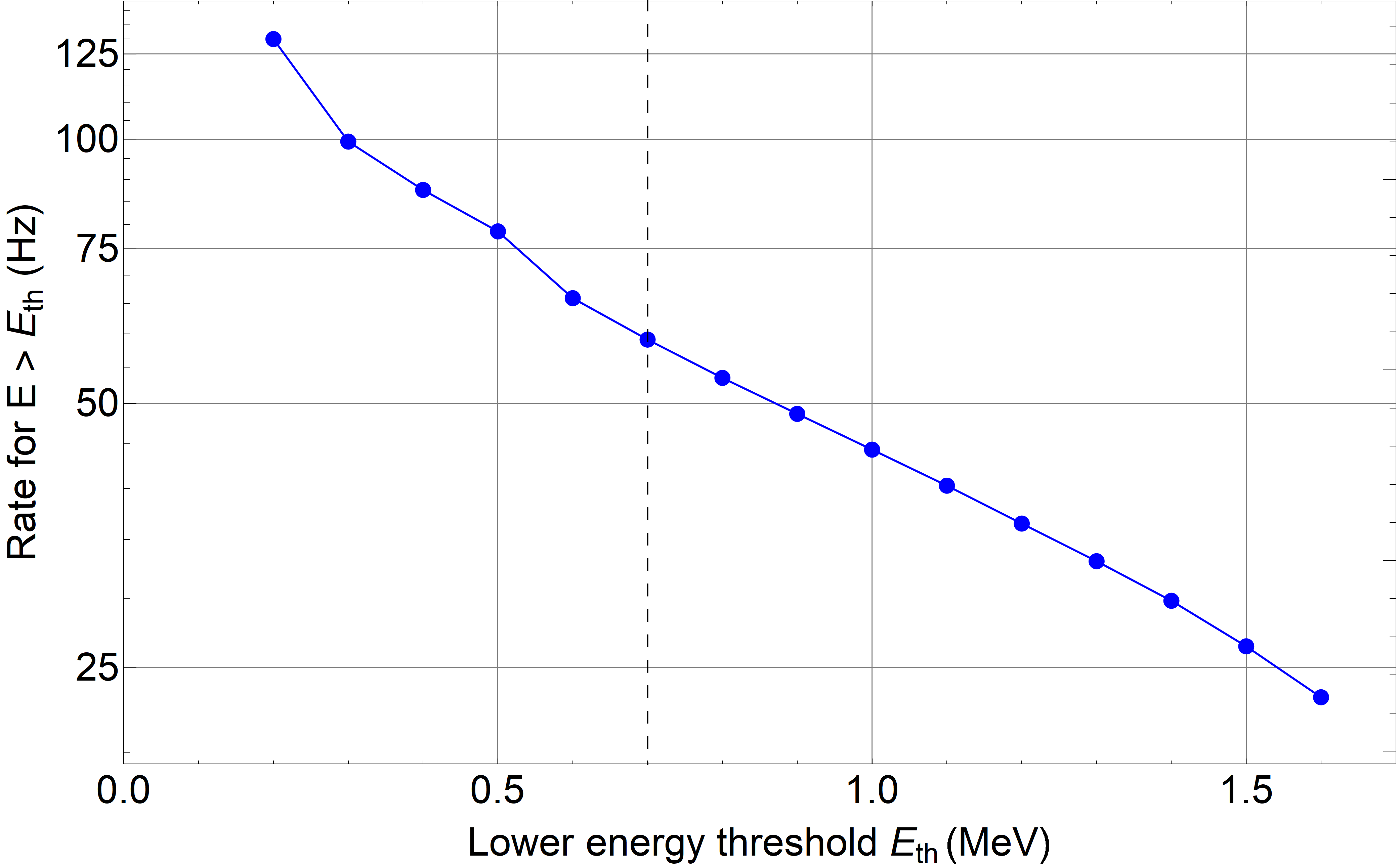}
            \caption[]{\label{fig:rate-threshold} }
        \end{subfigure}
        \begin{subfigure}{0.5\textwidth}
            \centering
            \includegraphics[width=\textwidth]{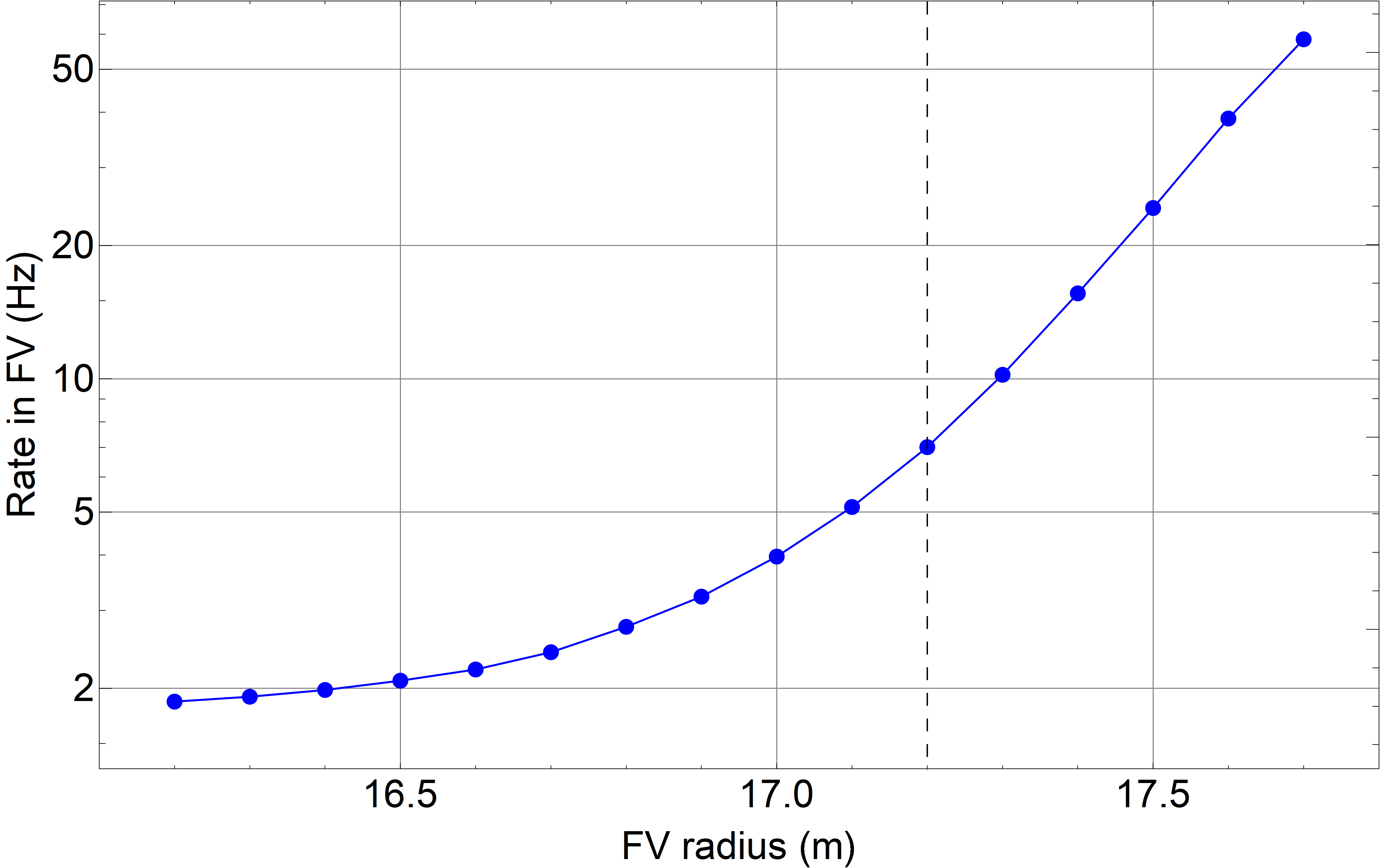}
            \caption[]{\label{fig:rate-fv}}
        \end{subfigure}
        \caption[]{Full background simulation results with SNiPER including all materials composing the detector with the contaminations listed in Table\,\ref{tab:impurities}. (a) Singles event rate as a function of energy threshold  in DV.  About 60~Hz total singles rate is achieved with 0.7 MeV energy threshold (shown as dotted line). (b) Singles event rate with energy larger than 0.7 MeV as a function of the FV cut. The singles rate can be reduced down to 7 Hz within a radius of 17.2 m (shown as dotted line).} 
        \label{spectra}
\end{figure}

The impurity of each material is taken from Table\,\ref{tab:impurities} and the final SNiPER simulation results, including all above mentioned effects, as a function of energy threshold  \Eth\ and FV radius are shown in Figure~\ref{spectra}. The total count rate expected for JUNO above \Eth~=~0.7~MeV is about 60~Hz as shown in Figure~\ref{fig:rate-threshold}. About 70\% of the alpha events in the LS are removed after applying the \Eth\ cut due to their lower quenched energy. Additional alpha event identification and rejection are foreseen with an efficiency larger than 99\% \cite{Keeffe2011} during data taking by using well-known pulse shape discrimination techniques. The total singles rate with energy larger than 0.7 MeV as a function of FV cut is illustrated in  Figure~\ref{fig:rate-fv}, where a sharp decrease of the rate from the edge of the LS to the  center is clearly visible. Thus, the external background can be effectively removed with  a suitable FV cut, as anticipated by Figure~\ref{fig:acrylic-FV-plot}. 

\begin{table}[tbp]
\centering
\begin{tabular}{|c|c|c|c|c|c|c|c|c|}
\hline
   \multirow{2}{*}{Material} & \multirow{2}{*}{Mass} & \multicolumn{5}{c|}{Target impurity concentration} & \multicolumn{2}{c|}{Singles} \\ 
   \cline{3-9}
& & $^{238}$U & $^{232}$Th & $^{40}$K & $^{210}$Pb 
& $^{60}$Co 
& DV & FV \\
& [t] & [ppb] & [ppb] & [ppb] & [ppb] 
& [mBq/kg] & [Hz] & [Hz]  \\ 
\hline
\hline
LS-reactor & 20000 & 10$^{-6}$ &  10$^{-6}$ & 10$^{-7}$ &  10$^{-13}$
& & 2.5 & 2.2  \\ \hline
Acrylic &  610 & 10$^{-3}$ & 10$^{-3}$ & 10$^{-3}$ & & & 8.4 & 0.4  \\ \hline
\multirow{2}{*}{SS structure} &  1000 & 1 & 3 & 0.2 & & 20 & \multirow{2}{*}{15.9}  & \multirow{2}{*}{1.1} \\ 
   & 65 & 0.2 & 0.6 &  0.02 && 1.5  && \\ \hline
\multirow{3}{*}{PMT glass} & 33.5 & 400  & 400 &  40 && & \multirow{3}{*}{26.2} & \multirow{3}{*}{2.8} \\
& 100.5 & 200  & 120 &  4 && &&\\
& 2.6 & 400  & 400 &  200 && &&\\
\hline
\multirow{2}{*}{PMT readout} & 125 &68 & 194 & 5 &&16& \multirow{2}{*}{3.4} & \multirow{2}{*}{0.4} \\
& 16.3 & 93 & 243 & 12 &&14&& \\
\hline
Other && &&&& & 2.5 & 0.3 \\ \hline
\hline
\multicolumn{7}{|c|}{Sum} & 59 &  7.2  \\ 
\hline
\end{tabular}
\caption{Final background budget for the main materials used in the JUNO detector with  reconstructed energy $E_{rec}$ larger than 0.7 MeV. The expected count rates are given both in the full DV (\Rlab\,=\,17.7\,m) and in the default FV (\Rlab\,=\,17.2\,m). The ``Other'' components include all materials that have relatively smaller contribution to the background (compare with Table\,\ref{tab:singles-noFV}), such as the calibration parts, the LPMT cover, the rock, and the radon in water. These results include energy resolution, optical propagation, charge reconstruction, and non-uniformity corrections. 
\label{tab:bkgBudget}}
\end{table}

Table\,\ref{tab:bkgBudget} summarizes the contaminations and count rates of each material. The mass and the putative contaminations for each material are copied here from Table\,\ref{tab:impurities}  for convenience. Count rates are reported in the case of both DV (\Rlab\,=\,17.7\,m) and the default FV cut (\Rlab\,=\,17.2\,m). Compared with Table~\ref{tab:singles-noFV}, the final count rates in DV are larger after applying the energy resolution, especially the worse energy resolution at the edge. Based on the results of Table\,\ref{tab:singles-noFV}, we list in Table\,\ref{tab:bkgBudget} only those items
giving the dominant contributions (above 0.3 Hz in the FV), while  materials with small contributions are grouped together in ``Other''. In the last row of the table, the expected sum count rates from all material contaminants are shown: the main contributions are due to the glass of the PMTs, the SS  structure, and the LS itself. It has to be stressed once more that the impurity concentrations of the various detector components of Table\,\ref{tab:impurities} are presumed on the basis of preliminary screening measurements or of literature values, and should be updated once the final numbers will be available.  In any case, the JUNO requirement for a background count rate lower than 10\,Hz for the IBD analysis channel seems 
within reach.   

 The informations contained in Figure~\ref{spectra} can be combined in a single plot, which better shows the correlations among the different parameters.
The two dimension distribution of singles rate as a function of  volume (bottom X-axis) or radius (top X-axis) and reconstructed energy E$_{rec}$ (Y-axis) is shown in Figure~\ref{sec5:singlesR3E}. The contribution of the LS is clearly identified by its uniform distribution in the volume up to $\sim$1 MeV. Most of the events come from the external background (i.e. from the materials outside the LS) and can be effectively removed by the FV cut at r$_{LS}$=17.2 m illustrated by the dashed line.

\subsection{Accidental coincidences from natural radioactivity}
\label{sec:bkg-impact}

%
%

\begin{figure}[h]
 \centering 
 \includegraphics[width=.7\textwidth]{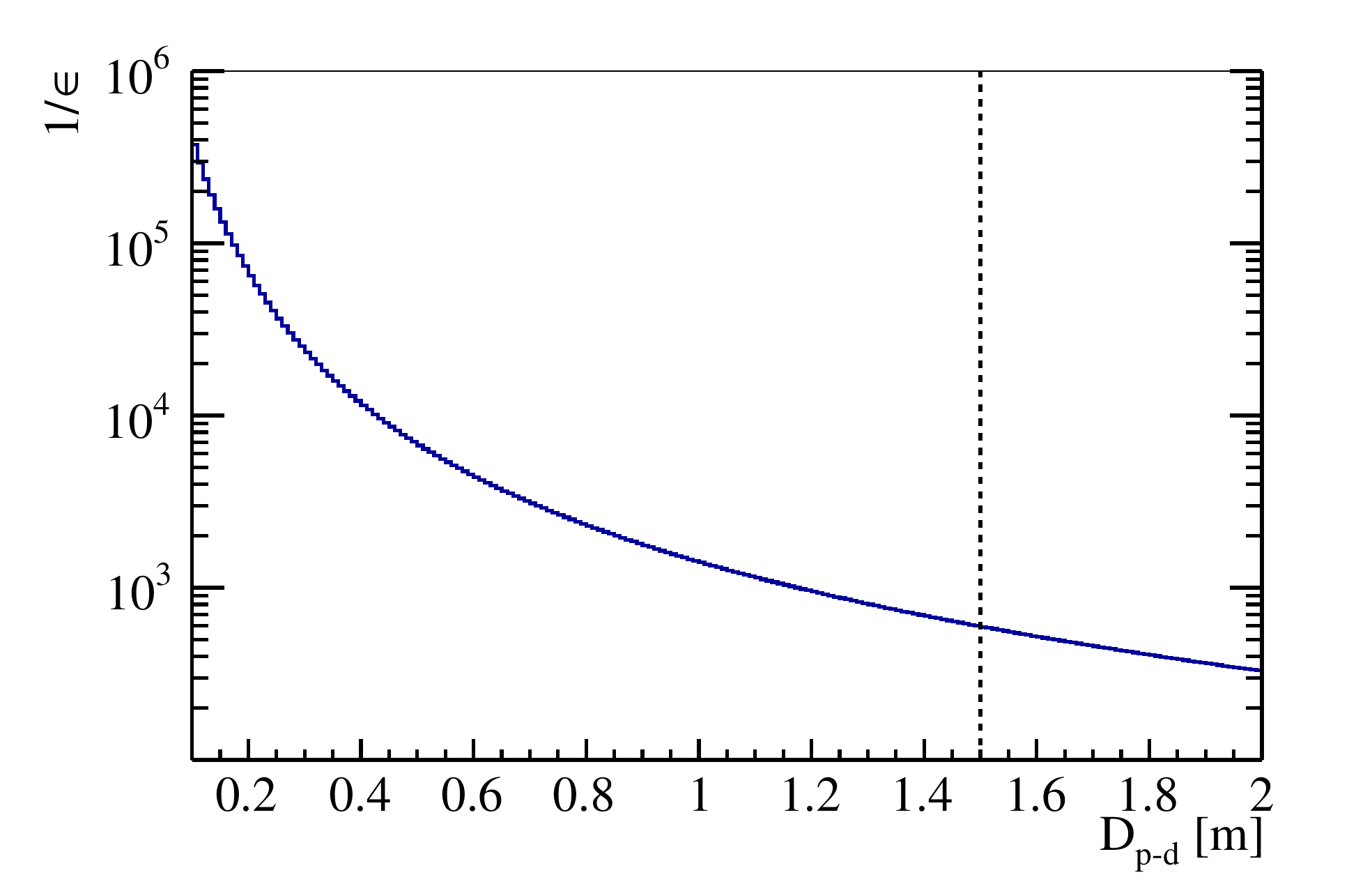}
  \hfill
  \caption{\label{sec5:accSupFactor}  The  integral distribution of the suppression factor (1/$\epsilon$) as a function of the distance \Dpd\ for a  time coincidence  window \DT~=~1\,ms between the prompt and the delayed signals with energy in the ranges $E_p$ = [0.7, 12]~MeV and $E_d$ = [1.9, 2.5]~MeV, respectively. The default \Dpd\  range (i.e. \Dpd~=~[0, 1.5]~m) is shown as the dashed line. }
 \end{figure}

As anticipated in Section~\ref{sec:nu-cuts}, we optimize the expected accidental coincidence rate $R_{acc}$ (see Equation~\ref{sec5:eq1}) with the help of a toy Monte Carlo and using the default selection cuts, as described in the following. One pair of events (prompt and delayed) in the chosen energy intervals ($E_p$ = [0.7, 12] MeV and $E_d$ = [1.9, 2.5] MeV) are randomly sampled 10$^9$ times from the distribution of Figure~\ref{sec5:singlesR3E}.
At each selected energy, the associated  frequency (rate) 
will depend on the position (radius) within the detector. The geometrical distance between each pair of  prompt and delayed events can be calculated by converting the spherical coordinates to polar coordinates. By choosing a time coincidence window between the prompt and delayed signals (\DT~<~1\,ms), related to the neutron lifetime in the IBD reaction, the  $R_p\cdot R_d\cdot\Delta T_{p-d}$ distribution as a function of the distance \Dpd\ can be calculated. Therefore, the selection of a specific \Dpd\  range of distances
between the prompt and the delayed events allows a further reduction of the accidental rate by the associated suppression factor (1/$\epsilon$), plotted in Figure\,\ref{sec5:accSupFactor}. The suppression factor decreases sharply at  large \Dpd\ distances, and this parameter can be tuned together with the time coincidence window \DT\ to optimize the IBD selection criteria according to the actual experimental conditions. 

Table~\ref{tab:opt} shows an example of  possible tuning of the resulting accidental rate $R_{acc}$ by different choices of the FV cut and of the threshold energy \Eth, with \DT\  window and \Dpd\  range fixed at 1\,ms and 1.5\,m, respectively. The final goal is to achieve the most favorable IBD signal to background ratio for the neutrino mass ordering analysis.
 \begin{figure}[h]
 \centering 
 \includegraphics[width=.7\textwidth]{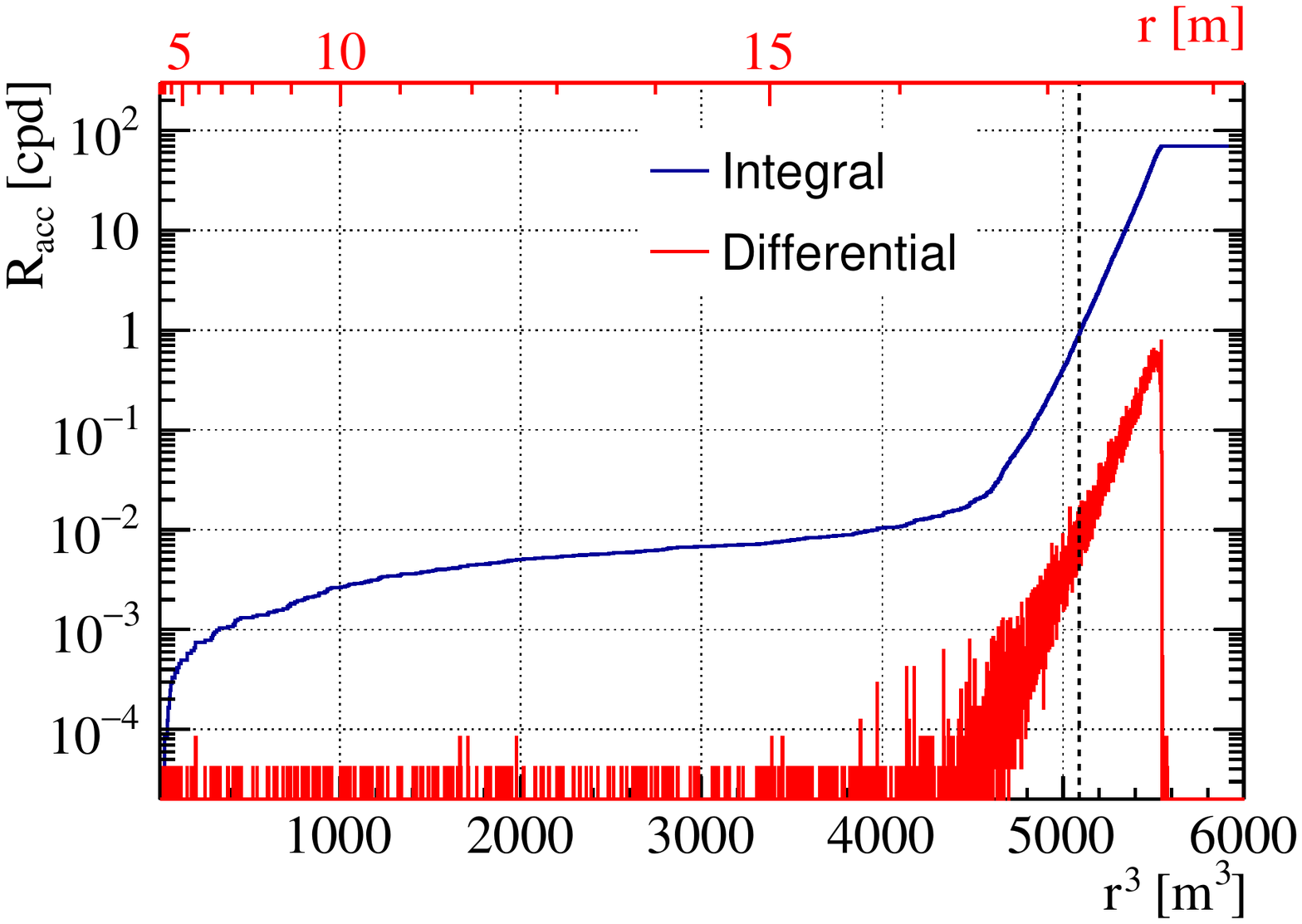}
  \hfill
  \caption{\label{sec5:accR3}  The differential (integral) distribution of the accidental rate as a function of radius with the default selection cuts of Section\,\ref{sec:nu-cuts} is shown as the red (blue) histogram. The default FV radius cut is shown as the dashed line.  }
 \end{figure}

The accidental rate  $R_{acc}$ as a function of  the volume (bottom X-axis) or radius (top X-axis) with default selection cuts and with the impurity inputs of Table~\ref{tab:impurities} is shown as the red histogram in Figure~\ref{sec5:accR3}, with the dominating contribution coming from the external contaminations.
The integral rate of the red histogram  (i.e. total $R_{acc}$ within the chosen FV) is shown as the blue histogram in Figure~\ref{sec5:accR3}.  
\begin{table}[tbp]
\centering
\begin{tabular}{|c|ccccc|}
\hline
   \multirow{2}{*}{R$_{acc}$ [cpd]} & \multicolumn{5}{c|}{Fiducial volume radius [m]}  \\ 
   \cline{2-6}
   & 17.0 & 17.1 & 17.2 & 17.3 & 17.4 \\ \hline
   $E_{th}$ = 0.7 MeV & 0.20 & 0.41 & 0.89 & 2.0 & 4.9  \\
   $E_{th}$ = 0.8 MeV & 0.19 & 0.38 & 0.83 & 1.9 & 4.6 \\
   $E_{th}$ = 0.9 MeV & 0.17 & 0.35 & 0.78 & 1.8 & 4.3 \\ \hline
\end{tabular}
\caption{ Evolution of the accidental rate $R_{acc}$ with FV and energy threshold cuts. An accidental background rate of 0.89 cpd is expected with default cut values \Eth~=~0.7\,MeV and FV with \Rlab\,=\,17.2\,m,  and with \DT\ coincindence window and \Dpd\ upper distance fixed at 1\,ms and 1.5\,m, respectively.
\label{tab:opt}}
\end{table}
The expected accidental background rate due to natural radioactivity is expected to be about 0.9~cpd with default cuts. As Table~\ref{tab:opt} shows, this rate is very sensitive to the selection of the FV radius with more than a factor 2 of increase or decrease when varying the FV cut by 0.1~m around 17.2~m. From Figure~\ref{sec5:accR3}, the accidental background is definitely negligible for $r <$ 16 m. Finally, with conservative assumptions, the accidental background rate from natural radioactivity is expected to be lower than the two other main backgrounds, i.e. the cosmogenic background ($\sim$1.6~cpd) and the geoneutrino background ($\sim$1.1~cpd), even without the muon veto \cite{ppnp}.

\section{Conclusion}
\label{sec:conclusions}
JUNO is a 20~kt LS detector whose primary goal is to determine the neutrino mass ordering by detecting reactor anti-neutrino from two nuclear power plants. With only 60~IBD events per day in the 0-10 MeV energy range, the control of the background sources is crucial. Apart from cosmogenic and geoneutrinos sources, the natural radioactivity of the materials is one of the key background with a deposited energy in the detector up to 5~MeV, overlapping with the anti-neutrino signal. This paper summarized the JUNO strategy in order to contain the count rate due to natural radioactivity at a level of 10~Hz in order to minimize its impact on neutrino mass ordering determination. 
Besides the optimization of the JUNO design, Monte Carlo  simulations and material screening  are being conducted to achieve the best radiopurity of the detector. Monte Carlo  simulations  were performed using three different codes (SNiPER, ARBY and G4-LA) in order to cross-check the results of the JUNO offline software, SNiPER, regarding the deposited energy and the induced singles rate in the full volume. An agreement within 2.5\% for the count rate  was achieved for LS and acrylic vessel  simulations, validating the physics inputs of SNiPER. After including the energy resolution, the optical propagation, the charge conversion, and the non-uniformity response of the detector in the full SNiPER simulation  (benefiting from the Daya Bay experience), the count rate for each component (LS, acrylic vessel, SS structure, PMT glass and readout, radon in water, calibration system, rock, etc.)  was derived assuming default cuts (0.7~MeV energy threshold and 17.2~m FV radius) and JUNO target radiopurity  requirements. These impurity concentration requirements, usually at the cutting-edge of available techniques, are based both on the experience of the previous experiments (especially for LS and radon in water), and on screening measurements performed on pre-production or production samples (acrylic, SS structure, PMT systems, calibration  parts) using several existing or developed on purpose screening facilities within the JUNO collaboration. The results of the full simulation showed that the total expected count rate in JUNO is $\sim$7~Hz, i.e. compliant with the 10~Hz target for the IBD analysis channel, with three main contributions from LS, SS structure, and PMT glass. Of course, the singles rate will be updated once the final numbers will be available. 
Based on a toy Monte Carlo using the count rate from natural radioactivity as well as time coincidence window of 1~ms and distance cut of 1.5~m for two singles events, an accidental background rate of $\sim$0.9~cpd  was derived. It demonstrates that this background rate is much lower than the IBD signal ($\sim$60~cpd) but also lower than the two other  main background sources, i.e. cosmogenics ($\sim$1.6~cpd) and geoneutrinos ($\sim$1.1~cpd). There  is room for improvement during the running phase of JUNO where the default cuts, including energy threshold, fiducial volume radius, time coincidence and distance cut, will be tuned according to the actual background count rate to optimize the signal over background ratio, in order to increase the exposure to the best of our ability for neutrino mass ordering determination.  The stricter requirements on the radiopurity of the LS (LS-solar) will make it possible to study solar neutrinos with the JUNO experiment, with the external background made totally negligible by the application of more stringent fiducial volume cuts \cite{JUNO-solar}.

\appendix
\section{SNiPER simulation framework}
\label{sec:sniper}

\begin{figure}[tbp]
\begin{subfigure}{.6\textwidth}
  \centering
  \includegraphics[width=.8\linewidth]{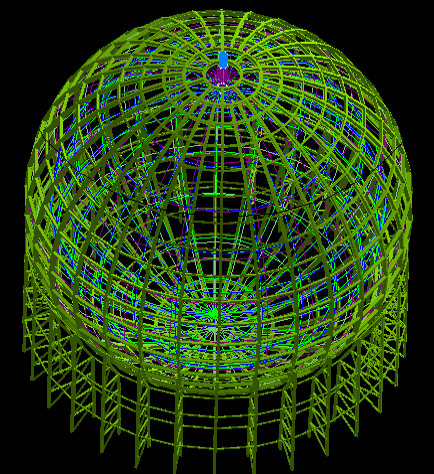}
  \caption{}
  \label{fig:sfig11}
\end{subfigure}%
\begin{subfigure}{.4\textwidth}
  \centering
  \includegraphics[width=.8\linewidth]{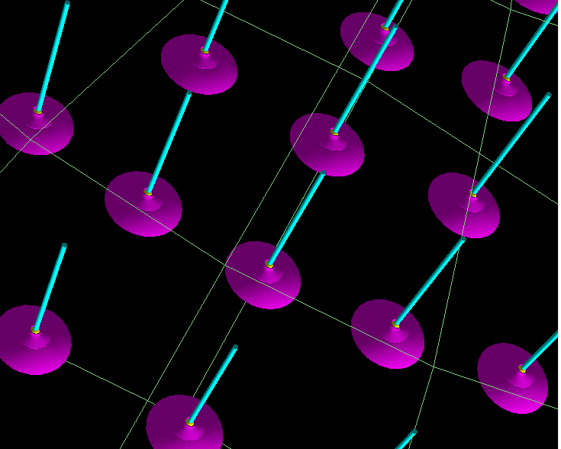}
  \caption{}
  \label{fig:sfig21}
\end{subfigure}%
\\
\begin{subfigure}{.4\textwidth}
  \centering
  \includegraphics[width=.8\linewidth]{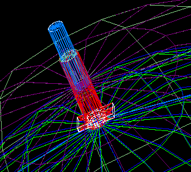}
  \caption{}
  \label{fig:sfig31}
\end{subfigure}
\begin{subfigure}{.6\textwidth}
  \centering
  \includegraphics[width=.8\linewidth]{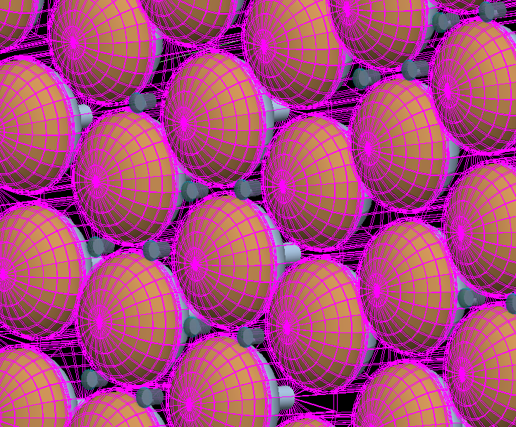}
  \caption{}
  \label{fig:sfig41}
\end{subfigure}
\caption{Details of the JUNO experimental setup as reconstructed by the SNiPER simulation code: a) SS truss supporting the PMTs and the acrylic sphere; b) Acrylic sphere containing the LS with bars on the outside for the connection to the steel truss; c) Chimney between the central detector and the calibration house on the top; d) Large and Small PMT systems.}
\label{fig:juno-by-sniper}
\end{figure}

The JUNO offline software is developed using the SNiPER (Software  for  Non-collider  Physics  ExpeRiments) tool~\cite{sec4:sniper1,sec4:sniper2}.  
The simulation framework is in charge of managing event data, detector geometries and materials \cite{kli_2018,szhang_2021}, physics processes, simulation truth information, etc. It glues physics generator, detector simulation and electronics simulation modules together to achieve a full simulation chain. 

For the Geant4-based detector simulation, the detector geometry includes all main materials, such as LS, acrylic sphere, SS structure, PMTs, water pool, and rock, and the visualization geometries \cite{You_2018,Zhu_2019} are shown in Figure~\ref{fig:juno-by-sniper}. 
The physics generators generate kinematic information of primary particles, which are saved into GenEvent objects. In the next step, the detector simulation algorithm accesses these GenEvent objects and starts tracking. Hits, which contain charge and time information, are generated in sensitive detectors and saved in SimEvent objects. The main parameters for PMT response are set to test data, which makes the results close to the real data in future. After that, the electronics simulation algorithm reads these SimEvent objects and performs the digitization, which generates ElecEvent objects containing waveforms information. These waveforms are processed by PMT calibration algorithm and CalibEvent objects are saved. The event reconstruction algorithm \cite{lizy_2021,qianz_2021,huang_2021} performs the event reconstruction by reading CalibEvent objects and stores RecEvent objects. At last, physicists can perform any physics analysis from RecEvent objects.

In this paper, only optical simulation is done considering the quenching effect, absorption, re-emission etc. Most of the optical parameters are got from Daya Bay, and scaled to JUNO expected number. In the case of radioactive decay chain simulations (e.g. \udto, \thdtd), each daughter nucleus is simulated separately without secondary decay. At the end, all events are added together in a single spectrum assuming the chain   is in secular equilibrium.

\section{ARBY simulation code}
\label{sec:arby}

\begin{figure}[tbp]
\begin{subfigure}{.7\textwidth}
  \centering
  \includegraphics[width=.8\linewidth]{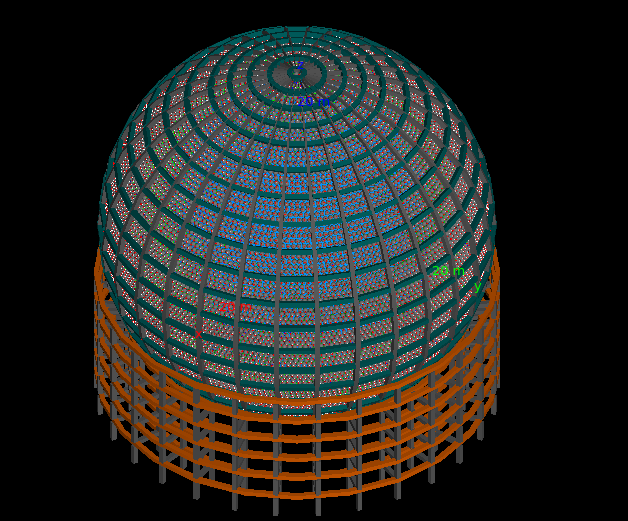}
  \caption{}
  \label{fig:sfig1}
\end{subfigure}%
\begin{subfigure}{.3\textwidth}
  \centering
  \includegraphics[width=.8\linewidth]{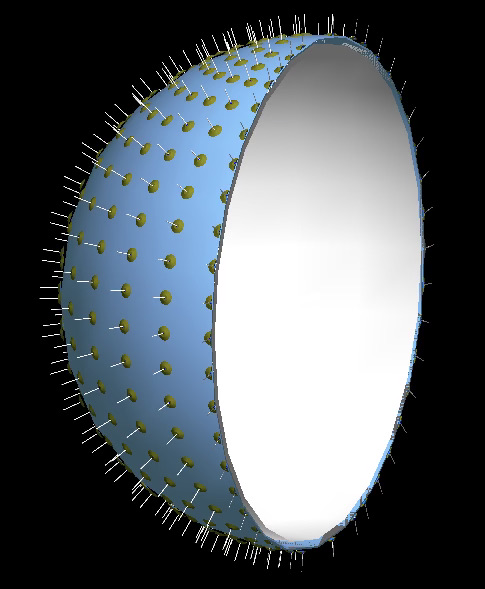}
  \caption{}
  \label{fig:sfig2}
\end{subfigure}%
\\
\begin{subfigure}{.3\textwidth}
  \centering
  \includegraphics[width=.8\linewidth]{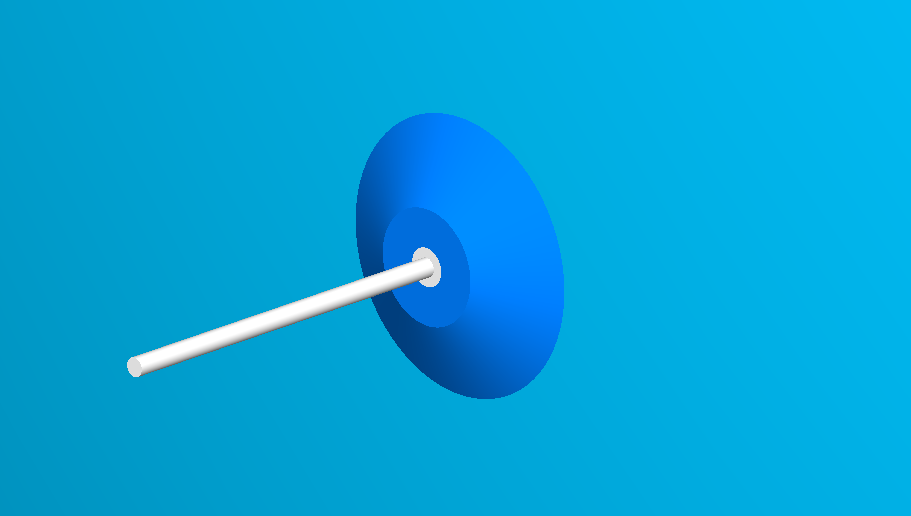}
  \caption{}
  \label{fig:sfig3}
\end{subfigure}
\begin{subfigure}{.35\textwidth}
  \centering
  \includegraphics[width=.8\linewidth]{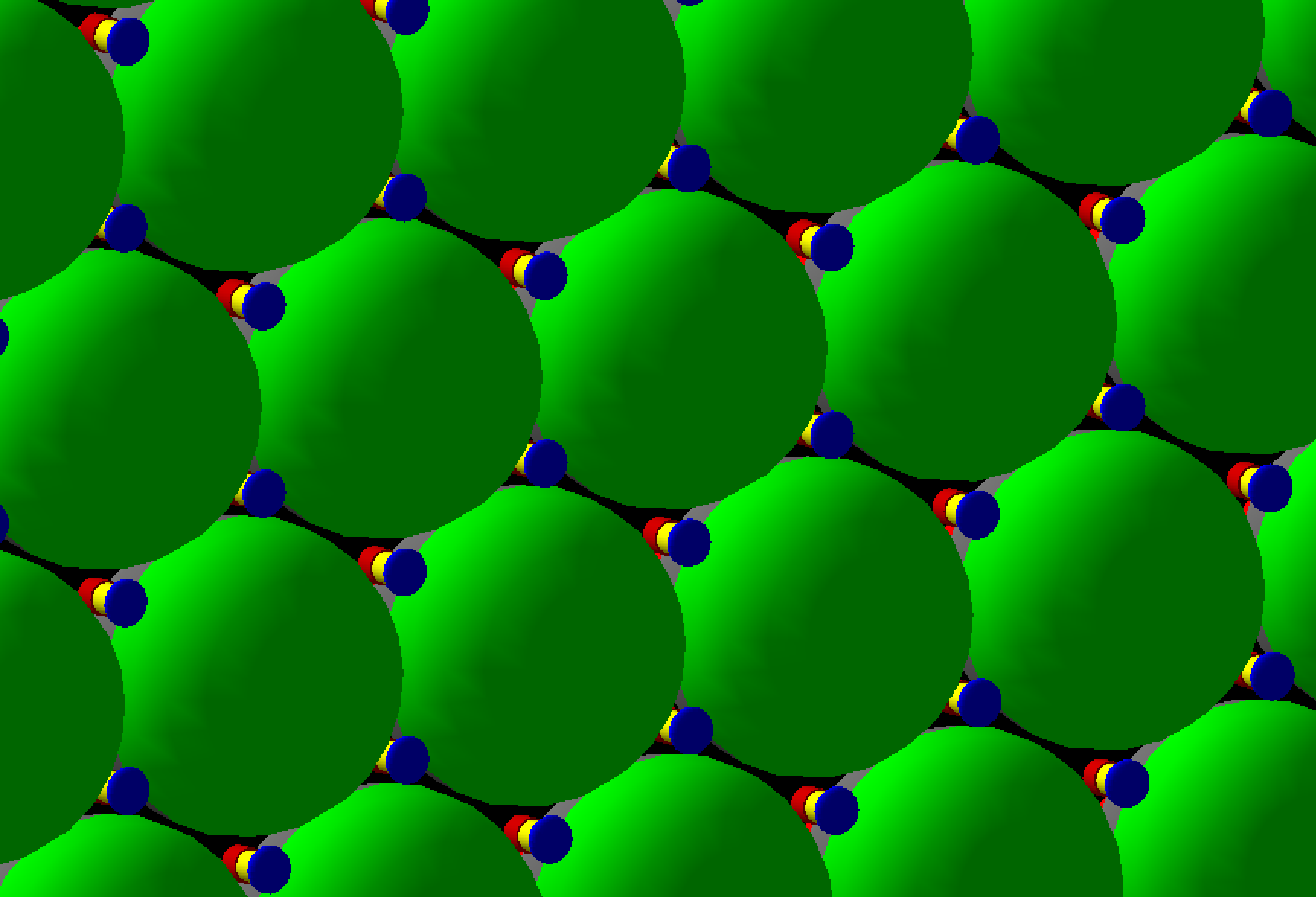}
  \caption{}
  \label{fig:sfig4}
\end{subfigure}
\begin{subfigure}{.3\textwidth}
  \centering
  \includegraphics[width=.75\linewidth]{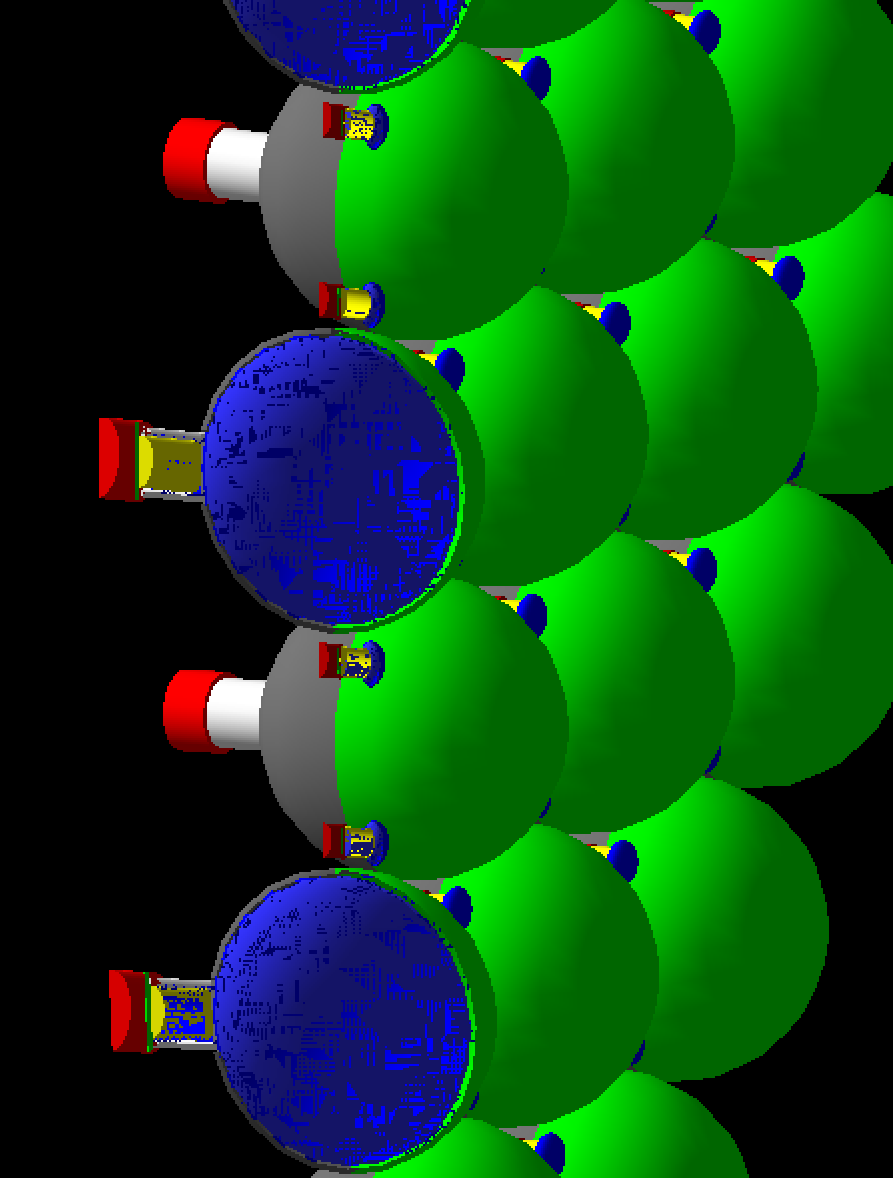}
  \caption{}
  \label{fig:sfig5}
\end{subfigure}
\caption{Details of the JUNO experimental setup as reconstructed by the ARBY simulation code: a) SS truss supporting the PMTs and the acrylic sphere; b) Acrylic sphere containing the LS with bars on the outside for the connection to the steel truss; c) Acrylic node and supporting SS bar; d) Large and Small PMT systems; e) Sections of the LPMTs showing potting elements.}
\label{fig:juno-by-arby}
\end{figure}

ARBY is a general purpose Monte Carlo simulation code — written by O.\,Cremonesi at INFN~Milano-Bicocca — designed for low energy particle physics applications. It is based on GEANT4 toolkit \cite{Geant4} and its main peculiarities are its high flexibility and ease of use. In ARBY, in fact, the geometrical description of the particular application to be simulated by the user (with very simple and intuitive commands) is completely separated from the physics implementation and code evolution, which are maintained by the developers. ARBY covers a wide variety of GEANT4 physics processes, includes the propagation of photons, alpha and beta particles, muons, neutrons and protons as well as nuclear recoils from decaying nuclides, and allows the generation of radioactive decay chains. For the latter, ARBY uses the G4RadioactiveDecay class of GEANT4, which offers the possibility of simulating the full disintegration of a nuclide through different processes profiting from an extensive data base (ENSDF \cite{ensdf}) to obtain all relevant information about the decay. When a daughter nucleus is itself unstable, like in the natural decay chains of \udto\ and \thdtd, ARBY allows to keep track of the single steps in the chain — i.e. of each individual decay — while scoring the event information. In this way the peculiarities of subsequent decays, like time correlations, do not get lost and allow to reconstruct particular effects which can manifest experimentally. Moreover, atomic effects (like X-ray production) and nuclear recoils accompanying each decay are also considered by ARBY, and included in the stored data. Other information like the position of the event in the detector volume or the particle quenching in organic scintillators can be also saved for each event.
The user needs just to describe the detector arrangement by means of a configuration file, where details concerning the materials in use as well as the position of the radioactive sources — i.e. the contaminant concentration of the different materials — are declared.

The ARBY software, and its subversions, are being used since many years by the INFN~Milano-Bicocca group for different applications, like the efficiency evaluation in gamma spectroscopy with HPGE diodes, or the investigation of the residual contamination sources and the development of accurate  background models in rare event experimental searches with bolometers (see e.g. \cite{cuore0-model,cuore-budget}).

In the case of JUNO, a configuration file was built to reconstruct with ARBY the main detector components: the LS, the acrylic sphere, the SS truss, the large and small PMT systems together with their readout electronics, the calibration equipment, the water pool, and the veto system. The different contamination sources were then systematically positioned in each of these materials and the resulting experimental spectrum evaluated with ARBY. Some details of the reconstructed geometries are shown in Figure\,\ref{fig:juno-by-arby}.

\section{G4-LA simulation code}
\label{sec:cenbg}

The G4-LA simulation code is based on GEANT4 toolkit using the G4RadioactiveDecay class in order to generate the decays of all the radionuclides of interest. The relevant contaminants (mostly U and Th chains and $^{40}$K) are generated uniformly in the bulk materials. The subsequent particles are tracked and the deposited energy as well as the coordinates of the mean deposited energy in the LS are recorded, allowing energy and FV cuts. The quenching effect for alpha and electrons particles has been taken into account as described in the paper. The detector geometry is simplified and includes the most critical materials of JUNO (LS, acrylic vessel and inner water pool). This allows a precise crosscheck of the energy spectra and the derived singles rate for LS and acrylic vessel in order to validate the Monte Carlo electromagnetic simulation of SNiPER.

\section*{Acknowledgements}

We are grateful for the ongoing cooperation from the China General Nuclear Power Group.
This work was supported by
the Chinese Academy of Sciences,
the National Key R\&D Program of China,
the CAS Center for Excellence in Particle Physics,
Wuyi University,
and the Tsung-Dao Lee Institute of Shanghai Jiao Tong University in China,
the Institut National de Physique Nucl\'eaire et de Physique de Particules (IN2P3) in France,
the Istituto Nazionale di Fisica Nucleare (INFN) in Italy,
the Italian-Chinese collaborative research program MAECI-NSFC,
the Fond de la Recherche Scientifique (F.R.S-FNRS) and FWO under the ``Excellence of Science – EOS” in Belgium,
the Conselho Nacional de Desenvolvimento Cient\'ifico e Tecnol\`ogico in Brazil,
the Agencia Nacional de Investigacion y Desarrollo in Chile,
the Charles University Research Centre and the Ministry of Education, Youth, and Sports in Czech Republic,
the Deutsche Forschungsgemeinschaft (DFG), the Helmholtz Association, and the Cluster of Excellence PRISMA+ in Germany,
the Joint Institute of Nuclear Research (JINR) and Lomonosov Moscow State University in Russia,
the joint Russian Science Foundation (RSF) and National Natural Science Foundation of China (NSFC) research program,
the MOST and MOE in Taiwan,
the Chulalongkorn University and Suranaree University of Technology in Thailand,
and the University of California at Irvine in USA.




\bibliographystyle{unsrt}
\bibliography{references}



\end{document}